\documentclass{article}
\pdfoutput=1
\usepackage{jheppub}

\usepackage{amsmath}
\usepackage{amssymb} 
\usepackage{empheq}
\usepackage{slashed}
\usepackage{amsfonts}
\usepackage{epsfig}
\usepackage{hyperref}

\newcommand{\be}{\begin{equation}}
\newcommand{\ee}{\end{equation}}
\newcommand{\ba}{\begin{array}}
\newcommand{\ea}{\end{array}}
\newcommand{\bea}{\begin{eqnarray}}
\newcommand{\eea}{\end{eqnarray}}
\newcommand{\sss}{\scriptscriptstyle}
\renewcommand{\S}{{\sss S}}
\newcommand{\T}{{\sss T}}
\renewcommand{\H}{{\sss H}}
\newcommand{\R}{{\sss R}}
\newcommand{\N}{{\sss N}}
\newcommand{\fN}{f_{\sss N}}
\renewcommand{\L}{{\sss L}}

\newcommand{\yl}{y_{\raisebox{-1pt}{\footnotesize $\ell$}}}

\begin{document}

\title{LHC constraints on dark matter with (130 GeV) gamma ray lines}

\author{James M.\ Cline, Grace Dupuis, Zuowei Liu}
\affiliation{Department of Physics, McGill University,
3600 Rue University, Montr\'eal, Qu\'ebec, Canada H3A 2T8}
\abstract{Dark matter annihilation into photons in our galaxy would 
constitute an exciting indirect signal of its existence, 
as underscored by tentative evidence for 130 GeV dark matter in
Fermi/LAT data.  Models that give a large annihilation cross section
into photons typically require the dark matter to couple to, or be 
composed of, new charged particles, that can be produced in colliders.
We consider the LHC constraints on some representative models of 
these types, including the signals of same-sign dileptons,
opposite-sign dileptons, events mimicking the production and 
decay of excited leptons, four-photon events, resonant production of
composite vectors decaying into two photons, and monophoton events.
}

\maketitle

\section{Introduction}
\label{section:intro}

The search for particle dark matter is being vigorously pursued from 
three complementary directions: direct detection by its scattering on
nucleons, indirect signals from annihilation in the galaxy, and its
production in colliders.  Ideally, a positive detection of dark matter
by one of these techniques would be corroborated by at least one of
the others.  Among the signatures amenable to indirect searches,  dark
matter annihilating into photons is interesting because of the
sensitivity of experiments like the Fermi Large Area Telescope and the
HESS II atmospheric Cherenkov telescope to the resulting gamma rays.
Indeed, there are hints that an excess of 130 GeV gamma rays possibly
due to such annihilations are coming from the galactic center
\cite{Bringmann:2012vr}-\cite{Fermi-LAT:2013uma}, and
even from external clusters of galaxies \cite{Hektor:2012kc} or from other 
sources within the galaxy \cite{Su:2012zg}.

  A relatively large cross
section $\sigma(\chi\chi\to\gamma\gamma) \sim 0.1\langle\sigma v\rangle_0$ is needed
to explain the observation \cite{Weniger:2012tx,Tempel:2012ey}, 
where $\langle\sigma v\rangle_0 \cong 1$pb$\cdot c$ is the
canonical annihilation cross section needed to explain the observed
relic density through thermal freeze-out.  For dark matter with mass
$\sim 100$ GeV, $\sim 0.1\langle\sigma v\rangle_0$ is also close to the upper limit
placed by Fermi/LAT in ref.\ \cite{Ackermann:2012qk}. 
Fermi itself \cite{Fermi-LAT:2013uma} finds a less significant excess
at 130 GeV than previous authors (local significance of 3.3$\sigma$), concluding that ``more data and
study are needed to clarify the origin of this feature.''  
Suggestions that the bump is due to instrumental noise have been 
studied \cite{Finkbeiner:2012ez,Hektor:2012ev}, with the conclusion
that such an origin is difficult to reconcile with the localization
of the signal near the galactic center.

Most particle physics models of dark matter do not predict such a
large value of $\sigma(\chi\chi\to\gamma\gamma)$, hence models that
do so  have been relatively less explored in terms of their
complementary predictions for the LHC; see refs.\
\cite{Bai:2012yq}-\cite{Domingo} for existing studies along
these lines.  In the present work we consider the implications for
LHC of two classes of models that have been proposed for the 130 GeV
line(s): one in which scalar dark matter $\chi$ couples to a new charge-2
scalar $S$, that mediates $\chi\chi\to\gamma\gamma$ through an $S$ loop 
\cite{Cline:2012nw,Buckley:2012ws}\footnote{for a similar model
in which vector dark matter couples to charged fermions, see
 \cite{D'Eramo:2012rr}.}  and
the other in which dark matter is a partially composite fermion with
a large magnetic moment, inherited from its charged constituents 
\cite{Cline:2012bz,Weiner:2012gm}.  
Both models involve new charged particles that transform under a hidden-sector
confining gauge group SU(2) or SU(3), and so they share some common
predictions, such as the production of $SS^*$ pairs that ``hadronize''
and decay into pairs of photons, leading to distinctive 4-photon
events.  Another common prediction is that exotic charged bound states
should be pair-produced  and decay into exotic final states,
either like-sign lepton pairs or lepton-photon pairs.  Neutral
bound states can also be produced singly as a resonance in the
$s$-channel analogous to $J/\psi$, with decays into fermion pairs.
There are the more familiar monophoton constraints coming from 
initial state radiation in the case where the resonantly produced
state decays into dark matter pairs.  In addition we find a novel
source of monophotons coming from the radiative decay of ``vector
mesons'' of the new SU(N) into their spin-0 ground states.

These classes of dark matter models that are capable of significant
$\chi\chi\to\gamma\gamma$ annihilation thus lead to a number of
low-background signatures for LHC, and we wish to characterize the 
extent to which LHC can constrain such models when the cross section
$\sigma(\chi\chi\to\gamma\gamma)$ is close to constraints from 
Fermi/LAT observations.  We start in section
\ref{section:models} with a review of the models considered, and
estimates of their respective cross sections for annihilation into
monoenergetic photons at the galactic center.  We compute the
cross section for LHC production of charged scalar pairs
as a function of energy in sect.\ \ref{section:lhcxsec}.  This is
followed by an analysis of LHC constraints on the models from the 
processes of decays of doubly-charged scalars into same-sign leptons 
(sect.\ \ref{section:sslepton}), direct production of vector
``mesons'' decaying to leptons (sect.\ \ref{section:oslepton}), 
decays of singly-charged composite fermions into lepton plus photon
(sect.\ \ref{section:excited}), decays of neutral composite states 
to diphotons and diphoton pairs (sect.\ \ref{section:4photon}), 
and monophoton events (sect.\ \ref{section:monophoton}).  We
synthesize these constraints to give an overview of the viability for
the models to explain 130 GeV gamma rays (as well a generalizations
to masses of other possible future DM candidates) in sect.\ \ref{viable}
and summarize our findings in sect.\ \ref{section:conclusion}.

\section{Theoretical models} 
\label{section:models}
In this section we summarize the three classes of models that motivated this study,
focusing on their predictions for $\gamma$-ray lines from dark matter annihilation.
All of them 
involve a new confining SU$(N_c)_d$ gauge group (with subscript $d$ for ``dark'')
and electrically charged scalar particles $S$ that transform in the fundamental of
SU$(N_c)_d$.  

\subsection{Loop-mediated model}
In the first class of models, the dark matter is assumed to be a
scalar, with coupling $(\lambda_{\S\chi}/2)\chi^2|S|^2$ to the
new charged scalar $S$.  The annihilation
$\chi\chi\to\gamma\gamma$ is mediated by a loop of $S$.  To get a large enough
cross section to be relevant for current  $\gamma$-ray observations, the
loop-suppression of the amplitude should be overcome by a somewhat large electric
charge $q_\S \ge 2$, and the color multiplicity $N_c$ of $S$.  In ref.\ 
\cite{Cline:2012nw}, the interaction potential between $\chi$, $S$ and the 
Higgs boson $H$ was considered to be
\be
	V_{\rm int} = {\lambda_{\S\chi}\over 2}\chi^2|S|^2 + \lambda_{\H\S} |H|^2|S|^2
	+ {\lambda_{\H\chi}\over 2} \chi^2|H|^2
\ee
It was shown that a cross section for $\chi\chi\to\gamma\gamma$ consistent with the
value determined in ref.\ \cite{Weniger:2012tx} for 130 GeV dark matter could be
obtained for parameter values $q_\S=2$, $\lambda_{\S\chi}=3$, $N_c=3$, $m_\S = 170$ GeV, 
for example.  More generally, one can express the cross section  
$\langle\sigma v\rangle_{\chi\chi\to\gamma\gamma}$ in terms of the mass ratio
$r=m_\S/m_\chi$ as
\be
	{\langle\sigma v\rangle_{\chi\chi\to\gamma\gamma} 
	\over 0.1\langle\sigma v\rangle_0}
	=
	0.44\, \left(q_\S\over 2\right)^4 \left(\lambda_{\S\chi}\over 3\right)^2 
	\left(N_c\over 3\right)^2
\left({m_\chi\over 130{\rm\ GeV}}\right)^{-2}
	r^{-4}f(r)
\label{svl}
\ee
where $f(r) = 9r^4(1-r^2(\sin^{-1}(1/r))^2)^2\to 1$ for large $r$ and is numerically fit by the formula
$f(r)\cong 1 + {0.4/(r-0.972)}$ which is good to 6\% for any value of
 $r>1.001$.  (We define $f$ in this way so that the $r$ dependence
in (\ref{svl}) is all transparently in the $r^{-4}$ factor for $r\gg 1$.)
The combination
$r^{-4} f(r)$ reaches its maximum value $\cong 19.4$ when $r=1$.
Recall that $\langle\sigma v\rangle_0 = 1$ pb$\cdot c$ is the nominal relic density 
cross section.  

In order to avoid the problem of charged relics (namely the ``baryon'' made from a bound
state of $N_c$ $S$ constituents) it is necessary to introduce a
neutral fundamental field $T$, which is also taken to be scalar.
If $T$ is lighter than $S$, then the decay $S\to T ee$ can be mediated
by the dimension-5 operator $\Lambda^{-1} ST^* \bar e_R^c e_R$.
In general one could have couplings to any right-handed leptons,
$\Lambda_{ij}^{-1} ST^* \bar l_{\R,i}^c l_{\R,j}$.  These couplings
are constrained by LHC searches for
like-sign lepton pairs, as we will discuss in section 
\ref{section:sslepton}.  Below the dark confinement scale $\Lambda_d$,
the dimension-5 operator will evolve to a renormalizable Yukawa
interaction $\eta_{\S\T}\bar e_R^c e_R$ with coupling of order
$\Lambda_d/\Lambda$.  The charged $\eta_{\S\T}$ will decay before
big bang nucleosynthesis (where its presence would change primordial
abundances) as long as $\Lambda \lesssim 10^{16.5}(\Lambda_d/100{\rm\
GeV})^{3/2}$ GeV.

\subsection{Magnetic dipole DM model}
\label{mddmm}
In the second class of models, dark matter is a mixture of an
elementary fermion and a composite one made from charged constituents,
that can give a large transition magnetic moment
$\frac12\mu_{12}\bar\chi_1\sigma_{\mu\nu}F^{\mu\nu}\chi_2$
between the dark
matter $\chi_1$ and an excited state $\chi_2$ \cite{Cline:2012bz}.
We refer to these as magnetic dark matter (MDM) models.    The charged constituents are a fermion
$\psi$ and boson $S$ that transform in the fundamental of the
SU$(N_c)_d$ gauge symmetry.  In the simplest case, $N_c=2$.  The
charges of $\psi$ and $S$ are constrained by the prohibition on
stable charged relics, which in the confining $N_c=2$ theory would be
the lightest of the ``baryonic'' bound states $SS$, $\psi\psi$ and $S^*\psi$ (we take
$S$ and $\psi$ to have opposite charges).  It is sufficient to
introduce a renormalizable operator 
$\epsilon_{ab}S^*_a\bar l_\R\psi_b$ that leads to mass mixing of 
$S^*\psi$ with the standard model lepton $l_\R$, hence to decays of
the would-be charged relic into $l_\R$ and photon or dark matter.
This shows that $S$ and $\psi$ should have charges $\pm
1/2$.\footnote{One can admit larger charges $\pm(n+1/2)$ by using
higher-dimensional operators $\epsilon_{ab}S^*_a\bar l_\R\psi_b
(\bar l_\R l_\R^c/\Lambda^3)^n$ to induce the charged relic decays.
Here we assume the simplest possibility $n=0$.}  Since we take them to
be singlets under SU(2)$_L$, they have hypercharge $\pm 1$.

\begin{figure}[t]
\begin{center}
\includegraphics[scale=0.4]{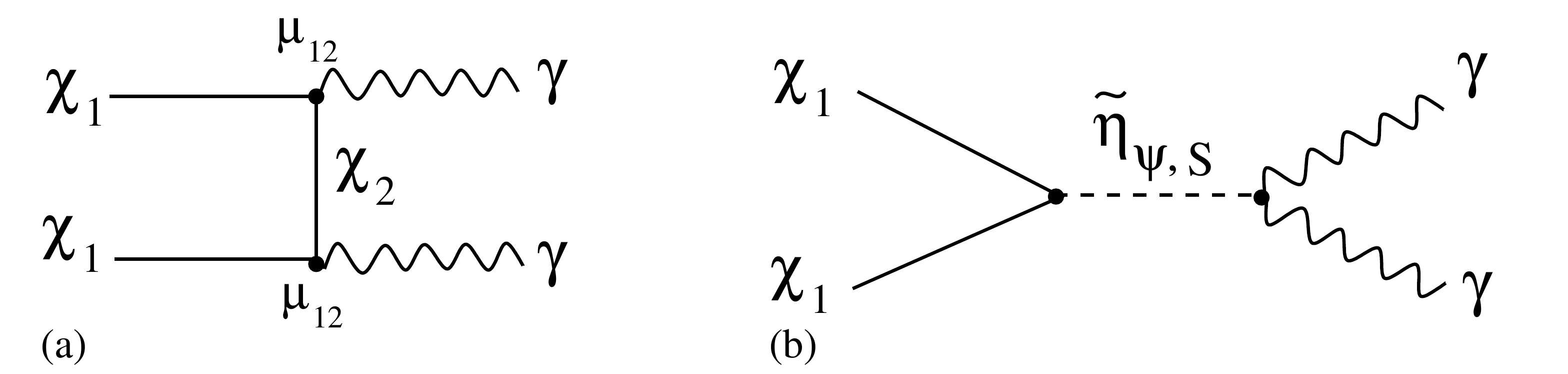}
\caption{Annihilation of dark matter into photons by (a)
magnetic moment interaction and (b) $s$-channel
exchange of the bound states $\eta_\S = S^*S$ or 
$\eta_\psi = \bar\psi\psi$.}
\label{fig:eta-chan}
\end{center}
\end{figure}

If the excited dark matter state $\chi_2$ is abundant in the
early universe, when DM annihilations are freezing out to fix the
relic abundance $n_\chi$, the magnetic-moment induced process 
$\chi_1\chi_2\to\gamma^*\to f\bar f$ (where $f$ represents standard model fermions) is so
efficient as to suppress $n_\chi$ below the value needed for $\chi_1$
to account for full dark matter density of the universe, leading to an
insufficient rate for $\chi_1\chi_1\to\gamma\gamma$ in the galaxy. This
can be avoided by arranging for $m_{\chi_2}-m_{\chi_1}\gtrsim 10$ GeV.
 In ref.\ \cite{Cline:2012bz} it was noted that for a
range of magnetic moment values $\mu_{12}$, 
annihilations $\chi_1\chi_1\to\gamma\gamma$, through the process of
fig.\ \ref{fig:eta-chan}(a), can have a cross section
that is larger than $\langle\sigma v\rangle_0$, so that $n_\chi$ is
again suppressed relative to the canonical value $n_0$, but that the
$\gamma$-ray signal in the galaxy is nevertheless at the observed
level.  Because $n_\chi$ scales as $1/\langle\sigma v\rangle$, we find
that the effective value of the cross section, as constrained by
searches for gamma ray lines, goes inversely to the actual cross
section:
\be
	\langle\sigma v\rangle_{\rm eff} \ =\   {
	\langle\sigma v\rangle_0^2\over 
	\langle\sigma v\rangle} \ \cong\ (0.04-0.1)\langle\sigma v\rangle_0
\ee
The range $0.04-0.1$ corresponds to the values estimated for the
tentative 130 GeV $\gamma$-ray signal by refs.\ \cite{Weniger:2012tx}
and \cite{Tempel:2012ey}.  In ref.\ \cite{Cline:2012bz} it was shown
that this range of cross sections corresponds to magnetic moments
in the interval $1.6 < \mu_{12}\cdot{\rm TeV}/f(r) < 2$ where
$r=m_{\chi_2}/m_{\chi_1}>1$ and $f(r) = \sqrt{(r^{-1}+r)/2}\ge 1$.
Note that $\langle\sigma v\rangle$ scales as $\mu_{12}^4$.

In this model, the dark matter gets its magnetic moment from the bound
state $\eta \equiv S\psi$, whose magnetic moment is estimated as that
of the fermionic constituent $\psi$, $\mu_\eta = e/(2m_\psi)$, in analogy to the
magnetic moments of baryons in the quark model.  The DM mass
eigenstates are mixtures of a Majorana fermion and the composite Dirac
state $\eta$, and there is some reduction of $\mu_{12}$ relative
to $\mu_\eta$ by a mixing angle $\theta$: 
$\mu_{12} = \cos\theta\,\mu_{\eta}$.  Ref.\ \cite{Cline:2012bz} (see
fig.\ 3) found
that 
$1/\sqrt{2} \le \cos\theta \le 1$ for the cases of interest.  Taking
$2m_\psi\sim 130$ GeV, this gives $\mu_{12}$ roughly consistent with
the desired range mentioned above.

Putting the above results together, we find that the effective
cross section as would be inferred by $\gamma$-ray observations is predicted to
be
\be
	\langle\sigma v\rangle_{\rm eff}  \cong 0.1\, 
	\langle\sigma v\rangle_0  \left({f(r)\over \cos\theta}\right)^4
	\left({m_\psi\over 100{\rm \ GeV}}\right)^4
\label{svm}
\ee
This prediction is only valid for
$\langle\sigma v\rangle_{\rm eff} \le \langle\sigma v\rangle_0$
since if this condition is violated, it means that the dark matter
density is larger than allowed by observations such those of 
\cite{Hinshaw:2012aka}.
The estimate (\ref{svm}) however assumes that there is no other
annihilation channel besides $\chi_1\chi_1\to\gamma\gamma$ mediated
by the magnetic moment interaction.  In fact, this model also has the
possibility of strong $\chi\chi\to\gamma\gamma$ annihilation through
the channel shown in fig.\ \ref{fig:eta-chan}(b).  This diagram must be
subdominant to that of fig.\ \ref{fig:eta-chan}(a) in order to justify
the estimate (\ref{svm}).  In ref.\ \cite{Cline:2012bz} it was argued
that this is true as long as the $s$-channel diagrams are not
resonantly enhanced.  But the other case is an interesting possibility
in itself, which can in fact also be incorporated in the loop-mediated
model.  We consider these $s$-channel models next.

\subsection{$s$-channel models}
\label{s-chan}

A third generic mechanism for producing $\gamma$ rays is for DM to
annihilate, possibly resonantly, into an intermediate particle $\eta$
that subsequently decays into two photons as shown in fig.\ 
\ref{fig:eta-chan}(b).  A natural realization of
this idea is for $\eta$ to be a bound state of a charged
particle-antiparticle pair, analogous to the $\eta$ meson of QCD; the decay
$\eta\to\gamma\gamma$ is then inevitable.  This scenario can be
achieved in either of the models presented above by an appropriate 
choice of parameters.

In the ``loop model,'' there is an enhancement in the limit $m_S\to
m_X$, which kinematically coincides with the picture where a bound
state $\eta_\S = S^*S$ is resonantly produced and {subsequently} decays to
photons.  In the strongly interacting description, we can identify
$S^*S = f_\eta \eta_\S$ where $f_\eta$ is a decay constant of order the
confinement scale, such that $\langle\eta_\S |S^*S|0\rangle = f_\eta$.  
The quartic coupling $\chi^2|S|^2$ then becomes a cubic interaction
$(\lambda_{\S\chi}/2) f_\eta \eta_\S \chi^2$.  By comparing to the theory of $\eta$ and $\pi^0$ decays in 
QCD, we can estimate the coupling of $\eta_\S$ to photons as
\be
	\alpha {q_\S^2 N_c\over 4\pi f_\eta}\,\eta_\S F_{\mu\nu}F^{\mu\nu}
\ee
where we have taken $\eta_\S$ to be even under parity and hence used
$F^2$ rather than $F\tilde F$.  
Although this result is adapted from the anomaly of the axial current
for fermionic constituents, we will assume that a similar result holds
for the present case of bosonic constituents.  The partial width for
$\eta_\S\to\gamma\gamma$ is then 
\be
	\Gamma_{\eta_\S\to\gamma\gamma} =  {\alpha^2\, N_c^2\, q_\S^4\over
	64\pi^3\, f_\eta^2} m_\eta^3
\ee
which is 4 MeV for the parameter choices $q_\S=2$, $N_c=3$, $m_\eta=260$ GeV, $f_\eta=130$ GeV for
example.  The width for $\eta_\S\to \chi\chi$ is 
\be
	\Gamma_{\eta\to\chi\chi} = {\lambda_{\S\chi}^2 f_\eta^2\over 16\pi\,
        m_\eta}\sqrt{1- 4m_\chi^2/m_{\eta_\S}^2}
\ee
This is generically much larger than
$\Gamma_{\eta\to\gamma\gamma}$, unless the former is kinematically 
forbidden.  The cross section for $\chi\chi\to\gamma\gamma$
corresponding to fig.\ \ref{fig:eta-chan}(b) is
\be
	\langle\sigma v\rangle_{\chi\chi\to\gamma\gamma}
	= {\alpha^2\lambda_{\S\chi}^2 N_c^2 q_\S^4 m_\chi^2\over 32\pi^3\left(
	(4m_\chi^2-m_\eta^2)^2 + \Gamma^2 m_\eta^2\right)}
\ee
Defining $\Delta r = m_{\eta_\S}^2/4m_\chi^2-1$ and assuming $|\Delta r|
\gg \Gamma m_{\eta_\S}/4 m_\chi^2$, we find that 
\be
	{\langle\sigma v\rangle_{\chi\chi\to\gamma\gamma}\over
	0.1 \langle\sigma v\rangle_0} = 
\left(q_\S\over 2\right)^4 \left(\lambda_{\S\chi}\over 3\right)^2 
	\left(N_c\over 3\right)^2
\left({m_\chi\over 130{\rm\ GeV}}\right)^{-2}{0.9\over (\Delta r)^2}
\label{svs}
\ee
This is roughly consistent with (\ref{svm}), showing that we can
obtain a similar estimate from the perturbative loop calculation as
from the bound state picture.  In both cases, the annihilation cross
section becomes enhanced relative to the generic value when
$m_\eta$ is close to $2 m_\chi$.  

For the MDM model of section \ref{mddmm}, the dark matter is fermionic,
and the interaction $g\bar \chi\eta_\S\chi$ has dimension 4; the  
coupling $\lambda_{\S\chi}$ in (\ref{svs}) should be replaced by
the Yukawa coupling $g$.  This interaction is generated by strong
dynamics, unsuppressed by any flavor symmetries, because the
$\chi$ states contain $S\psi$ or $S^*\bar\psi$ constituents, which
have a large overlap with $\eta_\S$ or $\eta_\psi$ via the
annihilation of the extra $\psi\bar\psi$ or $SS^*$ pair.  $g$ is therefore analogous
to the pion-nucleon coupling which is known to be large, $g\sim 14$
(see for example \cite{Ericson:2000md}).  
This can compensate the 
suppression of 256 relative to the loop model, from the smaller
charge $q_\S = 1/2$, and a further suppression of 2.25
from the smaller value of $N_c$.  To overcome these factors, one
needs to be somewhat close to resonance, with $\Delta r = 0.2$, thus
requiring a tuning of $m_\eta/2m_\chi$ at the level 
of 10\%.

\subsection{Dark glueballs}
\label{subsect:glueball}
In the following analysis, glueballs  of the hidden
SU(N)$_d$, which we denote by 
$\theta$, could play an important role because they might be the lightest
``hadrons'' of the hidden sector, if the confinement scale $\Lambda_d$
is lower than the mass of the colored constituents.  In QCD, there
exist candidate glueball states with mass around 1370 and 1700 {MeV}
\cite{Ochs:2013gi}.  In our models, 
the lightest glueball mass 
$m_\theta\sim \Lambda_d$ cannot be less than the dark matter mass;
otherwise the annihilation channel $\chi\chi\to\theta\theta$ would 
strongly suppress the DM relic density.

In these models, the main decay channel of the glueball
is into two photons, mediated by a loop of the charged constituent.
Thus the lightest glueball could be a Higgs boson imposter from the
perspective of the $h\to \gamma\gamma$ channel.  However if we take
$m_\theta$ to be greater than the constituent masses, the production
of glueballs from the decays of bound states will be kinematically
forbidden.  This removes one of the competing decay channels that 
would reduce the 
branching ratio of the bound states into two photons, which is of 
interest for constraining production of pairs of bound states, as
we discuss in sect.\ \ref{section:4photon}.

\section{LHC production}
\label{section:lhcxsec}

We begin our study of LHC constraints by computing the production
cross section for $pp\to S^*S$ ($\psi\bar\psi$), where $S$ ($\psi$)
is the new charged scalar (fermion)
that is taken to be neutral under SU(2)$_L$.
The relevant interaction Lagrangian, including the standard model
couplings to fermions, is given by 
\begin{equation}
\mathcal{L}_\text{int} = -ie q_\S (A_\mu-t_W Z_\mu) S^\dagger
\overleftrightarrow\partial^{\!\!\mu} S
-e q_f \bar{f} \gamma^\mu f A_\mu + e \bar{f} \gamma^\mu (\alpha_f + \beta_f \gamma_5)f Z _\mu 
\end{equation}
where $q_\S$ is the electric charge of $S$, 
$A\overleftrightarrow\partial_{\!\!\mu} B \equiv A\partial_\mu B - 
(\partial_\mu A) B$, 
$t_W \equiv \tan\theta_W$  ($\theta_W$ is the Weinberg angle), 
$q_f$ is the electric charge of fermion f, and 
$\alpha_f (\beta_f)$ is its vector (axial-vector) coupling. 
For up quarks, $\alpha_u = -5t_W/12 + c_W/4$,  $\beta_u =
-(t_W+c_W)/4$, and for down quarks, $\alpha_d = t_W/12 - c_W/4$, 
$\beta_d = (t_W+c_W)/4$, where $c_W\equiv \cos\theta_W$.

\begin{figure}[t]
\begin{center}
\centerline{\includegraphics[scale=0.33]{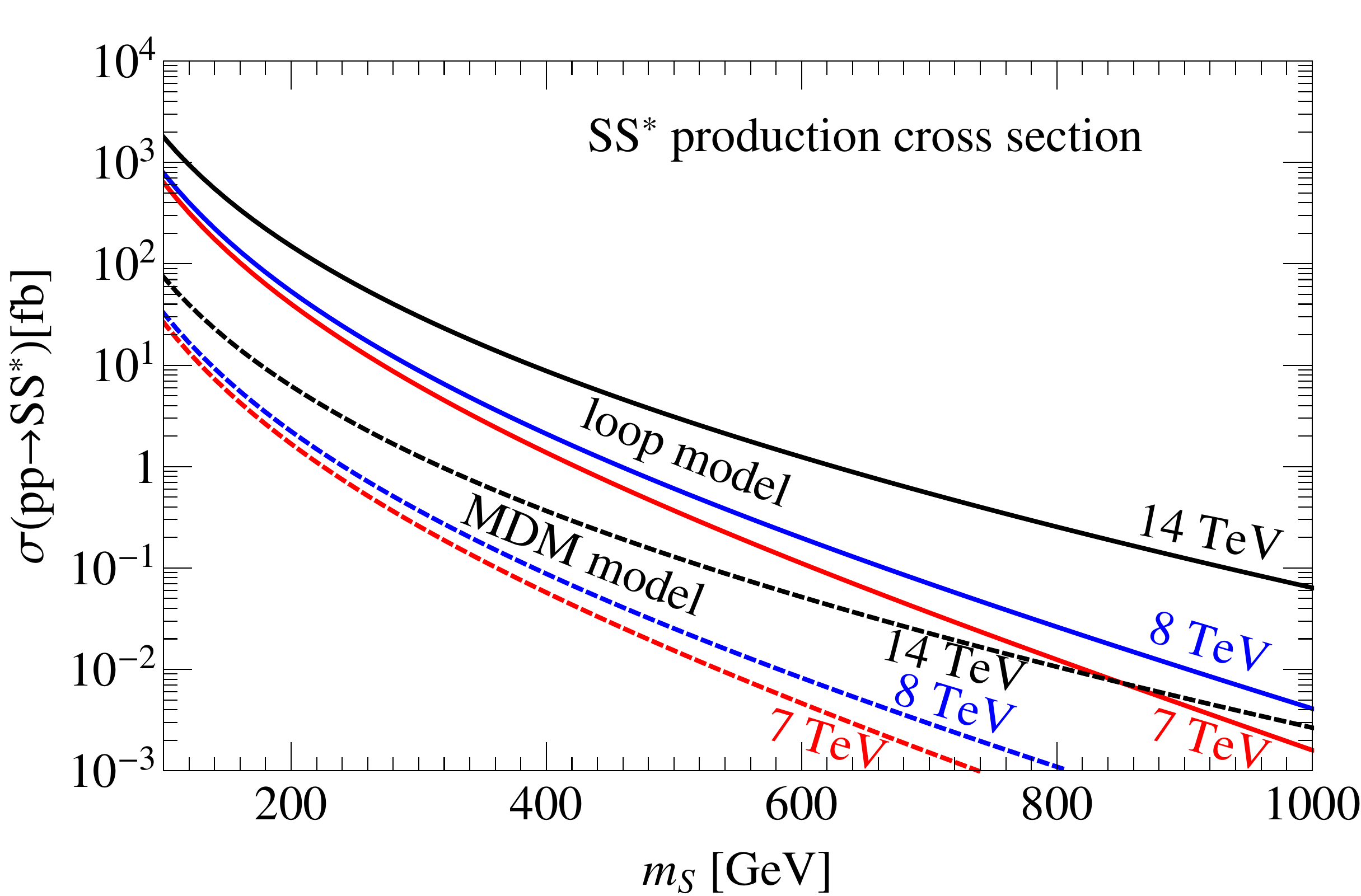}
\includegraphics[scale=0.33]{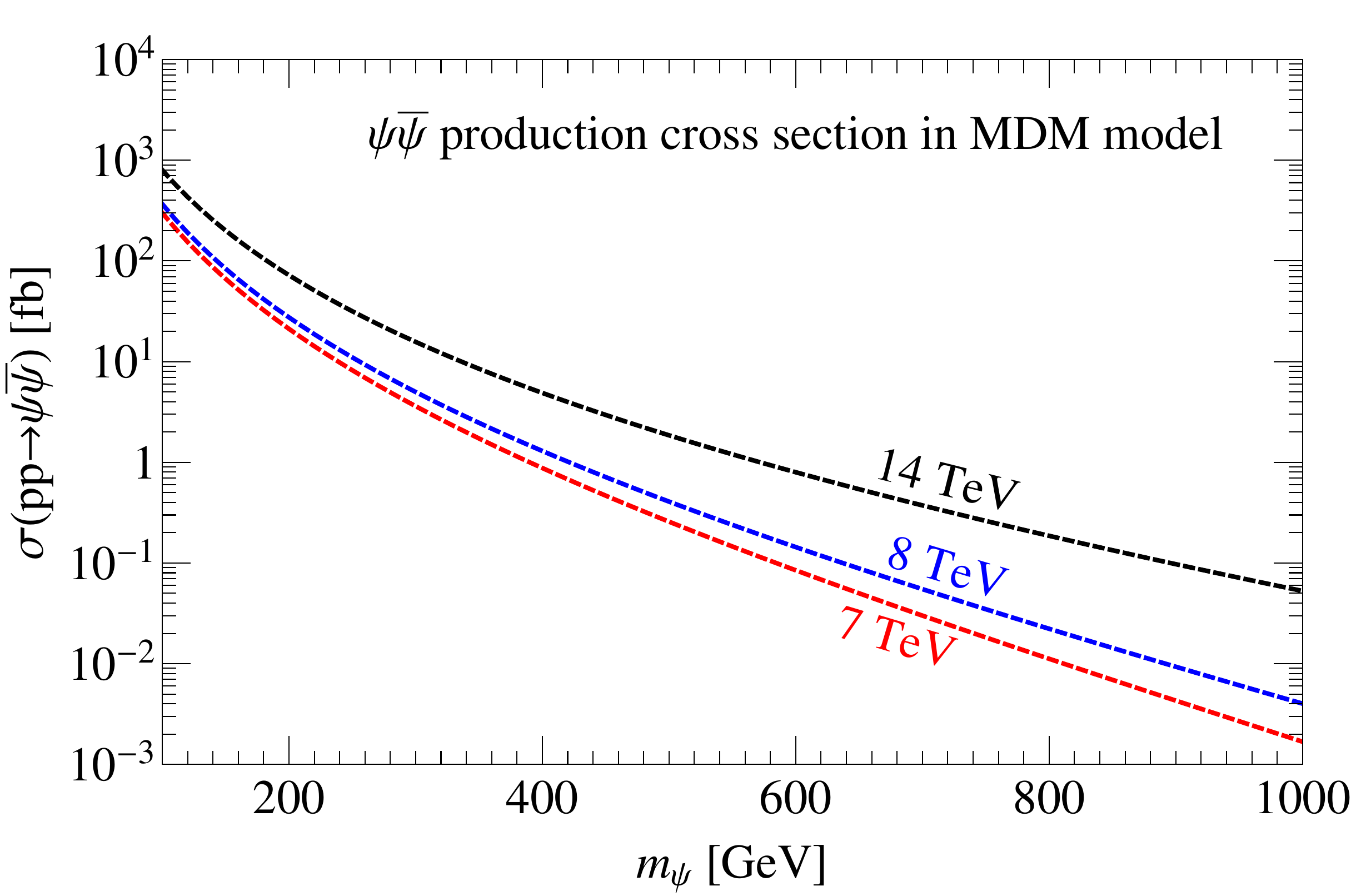}}
\caption{Left: the LHC production cross section of the  charged scalar pair 
as a function of the scalar mass, $m_S$, 
for three different collider energies: 7, 8, and 14 TeV.
Solid lines are for loop model with $q_s=2$, $N_c=3$, while dashed
ones are for magnetic model with $q_s=1/2$, $N_c=2$.
Right: same for production of fermion pairs $\psi\bar\psi$ in MDM
model.}
\label{fig:hadron_total}
\end{center}
\end{figure}

The parton level cross section for the process ($q\bar{q} \to \gamma/Z \to S\bar{S}$) is 
\begin{equation}
\begin{split}
& \hat\sigma (q\bar{q} \to S\bar{S}) = \frac{N_c}{N_q} \frac{\beta^3}{4} \left(\frac{4 \pi \alpha^2}{3 \hat{s}}\right) \\
& \times\left[ 
q_\S^2 q_q^2 
+ \frac{q_\S^2 t_W^2 (\alpha_q^2+\beta_q^2) \hat{s}^2}{(\hat{s}-m_Z^2)^2+m_Z^2 \Gamma_Z^2}
+  \frac{2 q_\S^2 t_W q_q \alpha_q (\hat{s}-m_Z^2)\hat{s}}{(\hat{s}-m_Z^2)^2+m_Z^2 \Gamma_Z^2}
\right]
\end{split}
\label{eq:parton}
\end{equation}
where $\hat{s}$ is the square of center-of-mass energy in the parton level, 
 $\beta = \sqrt{1-4 m_\S^2/\hat{s}}$, $N_q=3$ is the QCD color factor, 
$N_c$ is the hidden sector color factor of the $S$-scalar, 
$m_Z$ ($\Gamma_Z$) is the mass (decay width) of the $Z$ boson. 
For the analogous production of fermion pairs $\psi\bar\psi$, the
factor of $\beta^3/4$ in (\ref{eq:parton}) is replaced by $\beta(1 + 2
m_\S^2/\hat s)$.

The hadronic cross section ($pp \to \gamma/Z \to S\bar{S}$) at LHC is 
\begin{equation}
\frac{d\sigma(pp\to S\bar{S})}{dM_{S\bar{S}}} = K 
\frac{4M_{S\bar{S}}}{s} \sum_q \left.\hat\sigma_{q\bar{q}}
\right|_{\hat{s}=M_{S\bar{S}}^2} \int_\tau^1 \frac{dx}{x}\, q(x)\, \bar{q}(\tau/x)
\label{eq:hadron}
\end{equation}
where $\sqrt{s}=7, 8, 14$ TeV are the LHC energies considered for our
study, $\hat\sigma_{q\bar{q}} \equiv \hat\sigma(q\bar{q}\to S\bar{S})$ 
is the parton level cross section given in eq.\ (\ref{eq:parton}), 
$M_{S\bar{S}}$ is the invariant mass of the scalar pair $S\bar{S}$, 
$\tau=M_{S\bar{S}}^2/s$, and $q(x)$ and $\bar{q}(x)$ are the parton distribution functions of quark $q$ and $\bar{q}$ 
with momentum fraction $x$. Here we also use the $K$-factor to take into account the NLO QCD corrections. 
For our purpose, we take $K\simeq 1.3$ \cite{Muhlleitner:2003me}. 
Eq.\ (\ref{eq:hadron}) also 
applies for production of fermion pairs with the obvious substitution
$S\bar S\to \psi\bar\psi$.  

The total LHC production cross section for the scalar or fermion
pairs is plotted
in Fig.\ \ref{fig:hadron_total}.   Values of the charges $q_\S$,
$q_\psi$; and gauge group rank $N_c$ corresponding respectively to the 
loop and MDM models described above are adopted there.

\section{Same-sign dileptons}
\label{section:sslepton}

\begin{figure}[t]
\begin{center}
\includegraphics[scale=0.36]{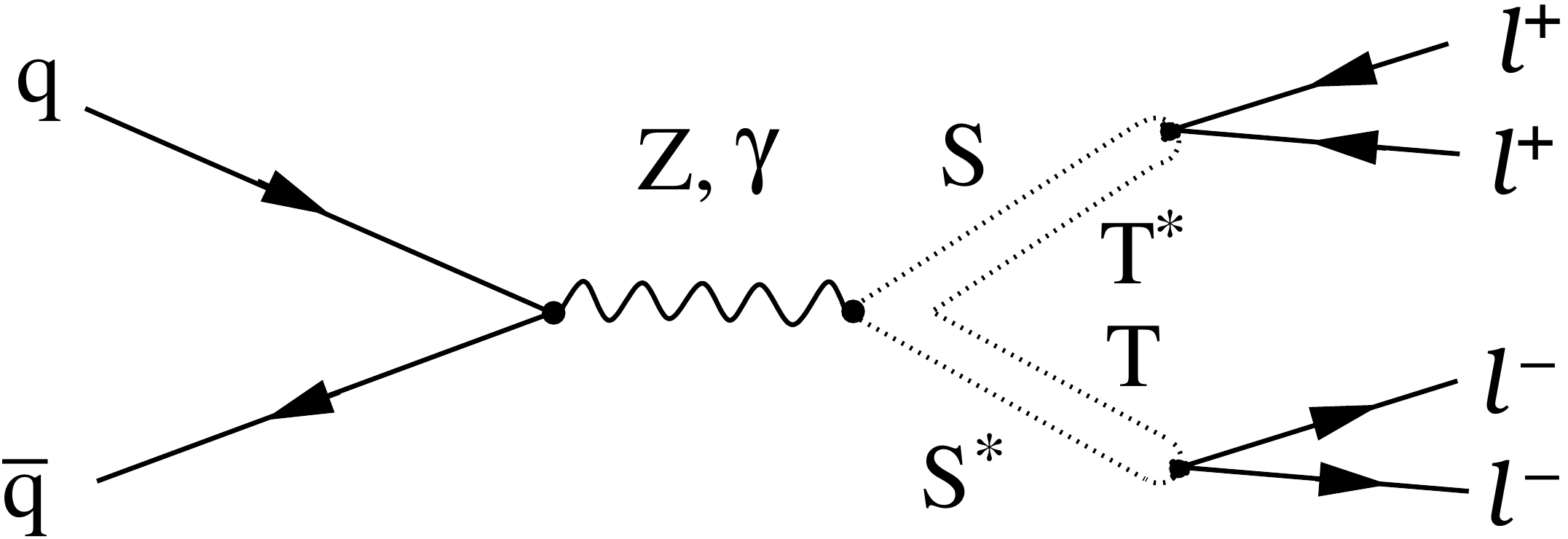}
\caption{Production of $\eta_{\S\T}$ pairs leading to like-sign dilepton events.}
\label{fig:ssdileptondiagram}
\end{center}
\end{figure}

Same-sign dilepton events can be produced at the LHC from the decay 
of the doubly charged scalars $S$ that are present in the loop model.
Recall that in order to avoid charged relics, this model introduces a dimension 5 operator  
$\Lambda_{ij}^{-1}T^*S\bar{\ell}_i^c\ell_j$ (where $i,j$ are
flavor indices of the right-handed leptons $\ell_i$) to permit such
decays, with $T$ being a lighter electrically neutral scalar, also
a fundamental of the SU(N)$_d$.
Since the SU(N)$_d$ interactions are confining, the $S\bar S$ pairs
produced as shown in fig.\ \ref{fig:ssdileptondiagram} will hadronize,
in this case into spin-0, charge-2 bound states of $S$ and $T^*$ which
we denote by $\eta_{ST}$.

\begin{figure}[b]
\begin{center}
\includegraphics[scale=0.5]{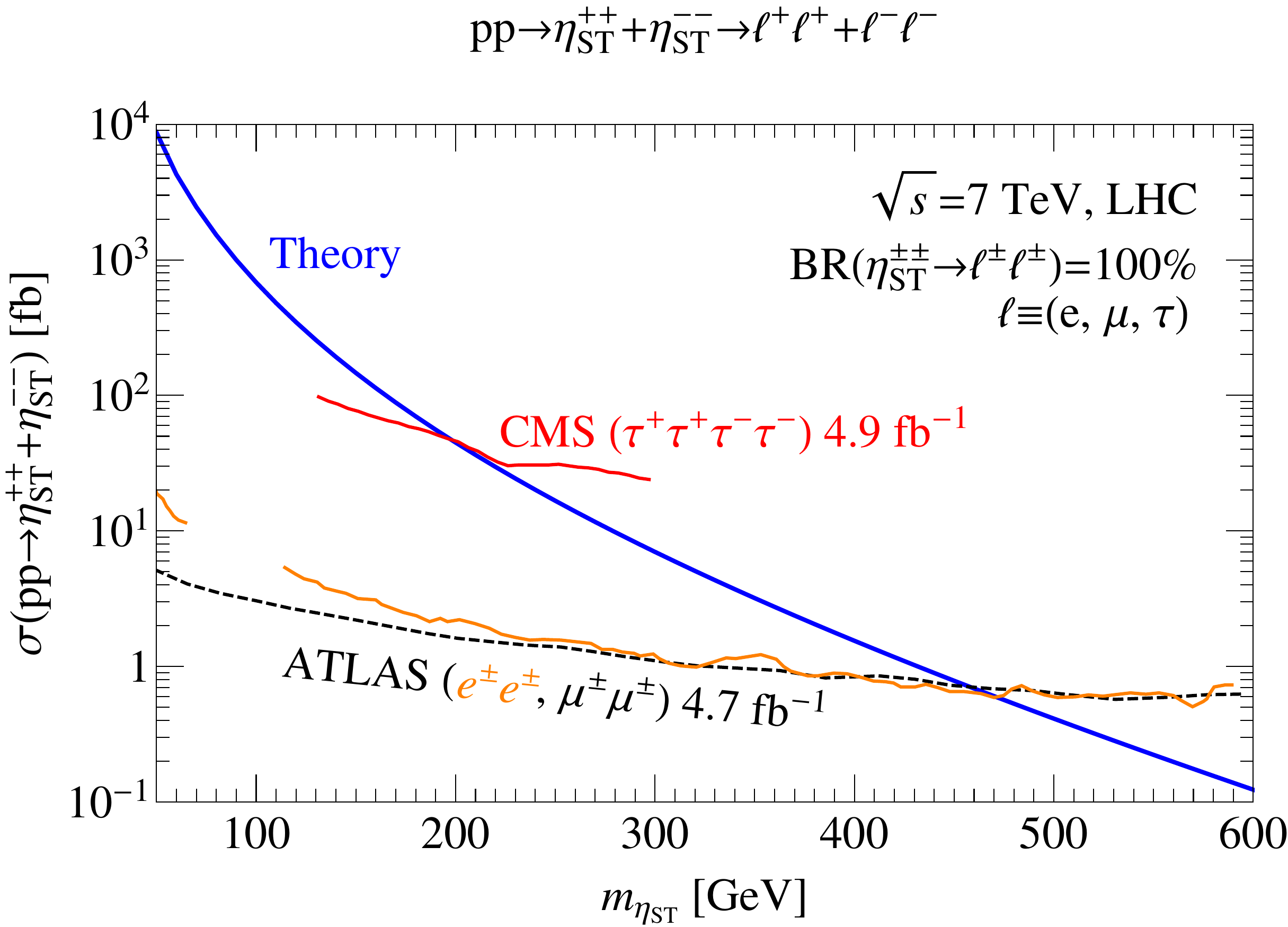}
\caption{Constraints on the production cross section for 
pairs of $\eta_{\S\T}$ mesons from same-sign dilepton searches of
CMS \cite{Chatrchyan:2012ya} and  ATLAS \cite{ATLAS:2012hi}, and the
prediction of the loop model. The mass range near $m_Z$ yields no
limits for the $ee$ channel due to the large background from
$Z \to e^+e^-$ with a charge misidentification.
}
\label{fig:ssdilepton}
\end{center}
\end{figure}

CMS and ATLAS have placed upper limits on the cross section for
production of like-sign dilepton final states
\cite{Chatrchyan:2012ya,ATLAS:2012hi}.  We have applied these to the
loop model and derived the  resulting constraints as a function of the
hidden meson mass $\eta_{ST}$, assuming that the dominant decays are
into  either $ee$, $\mu\mu$ or $\tau\tau$ final states.  The 
resulting exclusion curves are plotted in Fig.\ \ref{fig:ssdilepton}.
The model does not specify the flavor-dependence of the branching
ratios, but it may be natural to assume, in the spirit of minimal
flavor violation \cite{Ciuchini:1998xy}, that decays to $\tau\tau$
dominate.  These are also the least constrained because of the
difficulty of $\tau$ identification, but the mass range
$m_{\eta_{\S\T}} \sim 100-300$ GeV where the constraints have been
reported for this channel is quite relevant for our models, ruling
out $m_{\eta_{\S\T}} < 200$ GeV in this channel.  If $\eta_{\S\T}$
decays predominantly into $ee$ or $\mu\mu$, the more stringent constraint
$m_{\eta_{\S\T}}> 460$ GeV applies.

\section{Vector ``meson'' production}
\label{section:oslepton}

\begin{figure}[t]
\begin{center}
\includegraphics[scale=0.3]{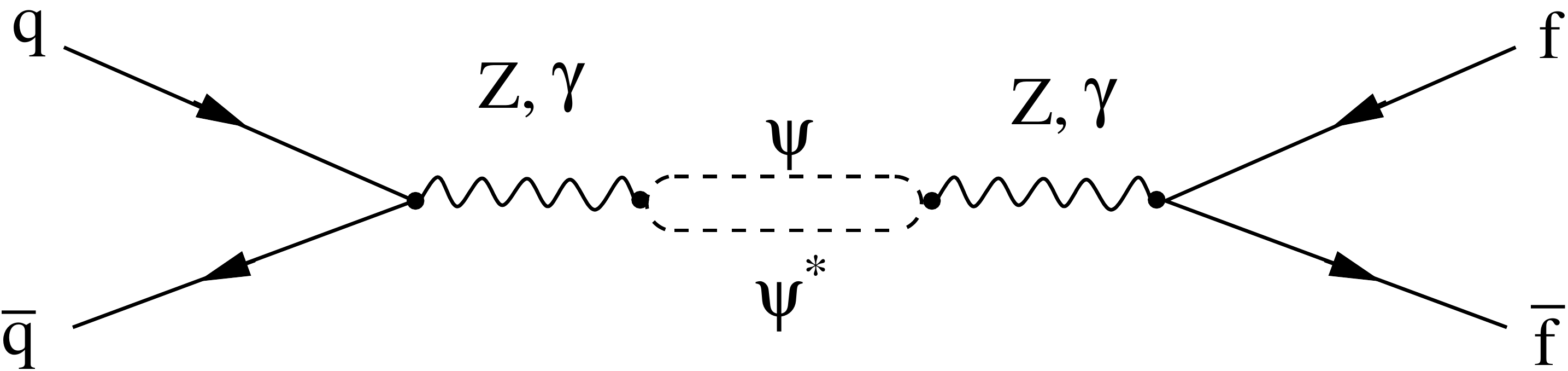}
\caption{Process that gives opposite-sign dilepton 
peaks in the s-channel resonance.  The bound state is a vector
``meson'' of the SU(2)$_d$ gauge sector, denoted by $\phi_\psi$.}
\label{fig:osdilepton_digram}
\end{center}
\end{figure}

If the pair $SS^*$ or $\psi\bar\psi$ is produced with net angular
momentum $J=1$, it can form a single vector meson bound state
analogous to the $\phi$ of QCD, rather than hadronizing into
two mesons, as shown in fig.\ \ref{fig:osdilepton_digram}.\footnote{We
 take $\phi$ or $J/\psi$ rather than $\Upsilon$
as the closer analogy because we are interested in the situation where
the constituent masses are below the confinement
scale, in order to avoid light glueballs, as discussed in section
\ref{subsect:glueball}.}\
  We refer to such states as $\phi_\S$ or $\phi_\psi$
in the models under consideration.  Once produced, the $\phi$
can decay into leptons or quarks.  Here we consider the decay into
leptons since it provides a lower-background signal that has been
searched for at the LHC.  The process can be viewed as a mixing of 
$\phi$ with the virtual photon or $Z$ boson, 
\begin{equation}
q\bar{q}  \to \gamma^*/Z \to \phi  \to \gamma^*/Z \to e^+e^-
\end{equation}
For center of mass energies much greater than $m_Z$, we can replace
the intermediate $\gamma$ and $Z$ with the weak hypercharge gauge
boson, allowing us to approximate  
$\Gamma(\phi \to \gamma^*/Z \to e^+e^-) = \Gamma(\phi \to
\gamma^*\to e^+e^-) / \cos^2\theta_W$.
The parton level production cross section is then
\begin{equation}
\sigma(q\bar{q} \to \phi \to e^+e^-) =  
{4 \pi\,q_q^2\over\cos^{4}(\theta_W)}\,\frac{\Gamma^2(\phi\to e^+e^-)}
{(\hat{s}-m^2_{\phi})^2+m_{\phi}^2\, \Gamma^2(\phi \to
\text{any})}
\end{equation} 
where the Breit-Wigner form of the resonance is assumed; here the
partial width refers to electromagnetic processes only, while the full
width includes the contribution from the $Z$.   The
observed signal is proportional to the area under the resonance curve, which
goes like $\Gamma(\phi\to e^+e^-)$ times the branching ratio for
$\phi\to e^+e^-$.

In the case of fermionic constituents, $\phi_\psi$ is just like the
$\phi$ of QCD, having orbital angular momentum $l=0$ and getting its
spin from that of the constituents.  Then the 
decay width of $\phi_\psi$ to electrons is  \cite{Peskin:1995ev}
\begin{equation}
\Gamma(\phi_\psi \to e^+e^-) = 
{4 \pi\, N_c\over 3} {\alpha^2 q_\psi^2\over E_\psi^2}\, |\Psi(0)|^2 
\label{gamma_phi_psi}
\end{equation}
where $\Psi(0)$ is the wave
function at the origin, $E_\psi = (p^2+m_\psi^2)^{1/2}$ is the kinetic
energy of the constituent, and $q_\psi = 1/2$ in the MDM model.  On the other hand, $\phi_\S$ must
have $l=1$ and this leads to the different result (see appendix
\ref{phiSamp})
\begin{equation}
\Gamma(\phi_\S \to e^+e^-) = 
{8 \pi N_c\over 3}{\alpha^2 q_S^2\over E_\S^2\,m_\phi^2}  |\vec{\nabla} \Psi(0)|^2
\label{gamma_phi_S}
\end{equation}

To estimate the wave function factors, we appeal to a semiclassical
model of mesons using a linear confining potential
with tension $k_d$, outlined in appendix \ref{bohr}.
In the nonrelativistic case, 
the size of the meson is expected to scale
as $(m k)^{-1/3}$ if $m$ is the mass of the constituents, and
therefore $|\Psi(0)|_{\phi_\psi}^2\sim m_\psi k_d$ where $k_d$ is the
string tension of the SU(N)$_d$ interaction.  Applying
eq.\ (\ref{gamma_phi_psi}) to the $J/\psi$ of QCD, one obtains
{$|\Psi(0)|_{J/\psi}^2 = (0.6\,{\rm fm})^{-3}$}.   Then we get the
estimate
\be
	{|\Psi(0)|_{\phi_\psi}^2\over 
	|\Psi(0)|_{J/\psi}^2 }  \cong {{m_\psi}\, k_d \over m_c\, k_{\sss
QCD}} \cong 10^6 
\label{psi0eq}
\ee
where we took $k_d/k_{\sss QCD} = 10^4$, corresponding to a
confinement scale of order 100 GeV, and $m_\psi/m_c=100$ since the
charm quark mass is $\sim 1.3$ GeV.  To estimate $\vec\nabla\Psi(0)$
for the $\phi_\S(1P)$ state, we take
\be
	|\vec\nabla\Psi(1P)| \sim |\vec p\,| |\Psi(1S)|
\label{gradpsi0eq}
\ee
where the momentum of the constituent is $p\sim (2\mu k_d)^{1/3} = 
(m_\S k_d)^{1/3}$ for the $n=2$ excited state $\phi_\S$.
We will also consider the relativistic regime in which $k_d\gg
m_\S^2$.  In that case the r.h.s.\ of (\ref{psi0eq}) is replaced by
$(k_d/k_{\sss QCD})^{3/2}$, and the estimate for $p$ in 
(\ref{gradpsi0eq}) becomes $p\sim (k)^{1/2}$.  One can use a
relativistic bound state from QCD instead of $J/\psi$ for the comparison of (\ref{psi0eq}) 
in that case; the $\phi$ of QCD is estimated to have $|\psi(0)|^2 =
(1.6\,{\rm fm})^{-3}.$\footnote{We correct the result given in
\cite{Peskin:1995ev} by taking into account the relativistic energy of
the constituent rather than just its mass in eq.\ 
(\ref{gamma_phi_psi}))}\ \ To interpolate between the relativistic and
nonrelativistic regimes, we fit the QCD wave functions to the ansatz
$|\psi(0)|^2 = a k^{3/2} + b m k$, taking $a=0.022$ and $b=0.13$.

\begin{figure}[t]
\begin{center}
\includegraphics[scale=0.5]{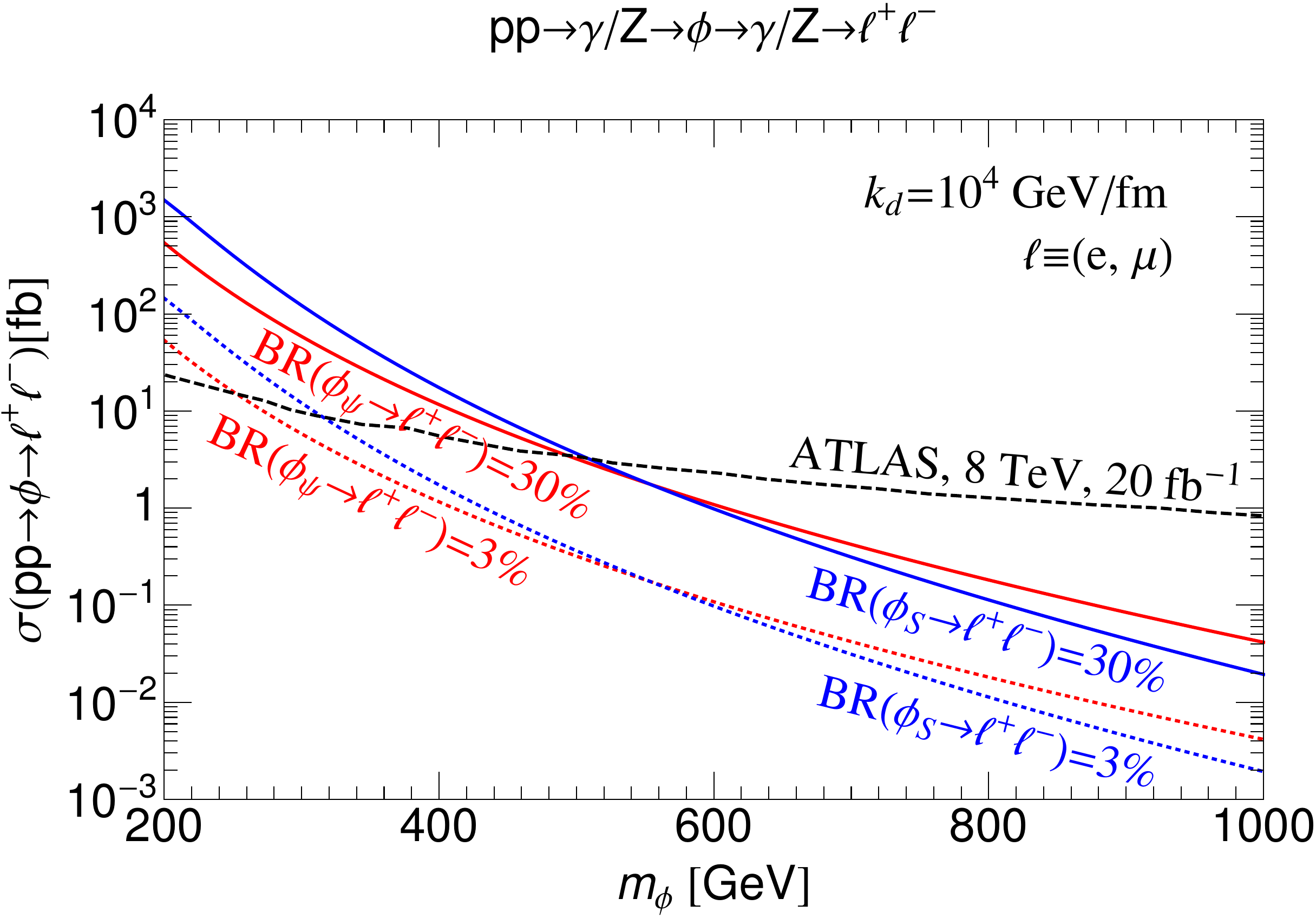}
\caption{{LHC constraints \cite{ATLAS:2013jma} on opposite sign 
same flavor dilepton resonances. 
We take $3\%$ or $30\%$ for the branching ratios into leptons 
$\ell^+\ell^- = e^+e^-+\mu^+\mu^-$. 
}}
\label{fig:dilepton}
\end{center}
\end{figure}

We applied LHC constraints on decays to $e^+e^-$ plus $\mu^+\mu^-$ final
states to our models, following
the recent ATLAS analysis \cite{ATLAS:2013jma} 
which utilized approximately $20$ fb$^{-1}$
integrated luminosity of data at 8 TeV center-of-mass energy, 
and also the recent CMS analysis \cite{CMS:dilepton} 
with the dielectron (dimuon) events sample 
corresponding to $19.6$ ($20.6$) fb$^{-1}$ integrated luminosity. 
Since the event samples from both ATLAS and CMS are comparable in 
size and the limits on the sequential standard model $Z'$ are 
also similar, we compare the vector meson production with 
only the expected 95\% C.L.\ limits on 
$\sigma(\phi)\times \text{BR}(\phi\to \ell^+\ell^-)$ from the ATLAS analysis. 
Fig.\ \ref{fig:dilepton} shows the theory predictions for
these final states  from the $\phi_\S$ and $\phi_\psi$ resonances,
assuming $q_\S=2$, $q_\psi=1/2$, and leptonic branching ratios of 3\%
or 30\%.  
The maximum possible branching ratio is determined by the 
available fermionic standard  model states:
$\text{BR} = 1/(3 + 1 + 8/3) = 0.15$ for a single lepton flavor.
We will argue in
section \ref{viable} that the true value in the loop or MDM models 
should be closer to 3\% due to competing decay channels that are lower
order in $\alpha.$

From fig.\ \ref{fig:dilepton} we infer the bounds
{$m_{\phi_\S}, m_{\phi_\psi} > 500$} 
GeV if the branching ratio for
$\phi$ decays into leptons is 30\%.  This limit goes down to 
{310 GeV and 250 GeV,} 
respectively for $m_{\phi_\S}$ and $m_{\phi_\psi}$, when
the branching ratio is 3\%.  In sect.\ \ref{viable} we will argue
that these limits are not difficult to satisfy in the dark matter
models since we have the freedom {to} make the confinement scale of the
SU(N)$_d$ interactions sufficiently large.


\begin{figure}[t]
\begin{center}
\includegraphics[scale=0.3]{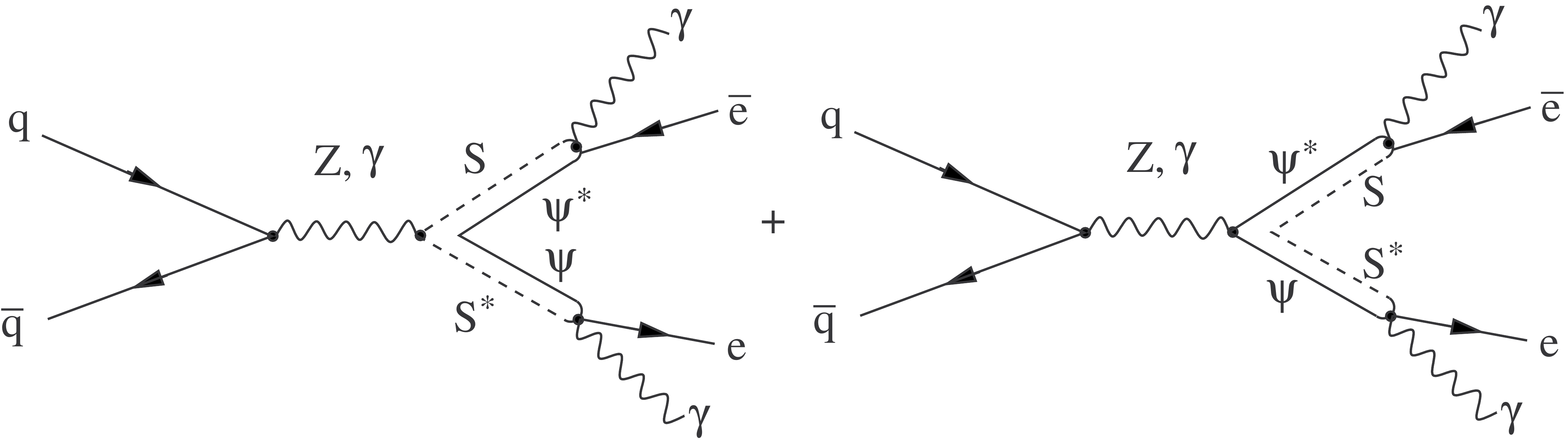}
\caption{Feynman diagrams for production of $N^\pm$ mesons that mimic excited electrons
decaying to $e^\pm\gamma$.}
\label{fig:excited_electron_diagram}
\end{center}
\end{figure}

\section{Excited electron/muon limits}
\label{section:excited}

In the MDM model, there are four possible hadronization processes that the scalar particle
$S$ can  undergo to form different bound states after being 
pair-produced.\footnote{Because we have assumed the gauge group is
SU(2)$_d$, $S$ can bind with any of $S$, $S^*$, $\psi$ or $\bar\psi$.}  One of
these results in pairs of charged particles 
 $N^- = S^*\psi$ and $N^+ = S\bar\psi$ with the same
quantum numbers as the right-handed electron and positron.
$N^-$ in fact mixes with
$e_\R$ and higher generation leptons through the Yukawa interaction
$y_i\epsilon_{ab}S^*_a\bar l_{\R,i}\psi_b$, where $i$ is the
generation index.  In the low-energy theory below the confinement
scale, this operator induces off-diagonal mass terms $m_i \bar 
l_{\R,i}N^-$, where $m_i = y_i\langle 0|\epsilon_{ab}S^*_a\psi_b|N^-\rangle$.
Couplings such as {$\bar{N}\slashed{A}e$} which were absent in the 
original flavor basis are not induced by diagonalizing the mass
matrix.   However, since
$N^-$ has a magnetic moment that is much smaller than that of the leptons
($\psi$ being heavy), a transition magnetic moment is induced between
the leptonic and exotic mass eigenstates.   Thus the decay 
$N^-\to e + \gamma$ can occur, which 
mimics the excited electron search signals predicted for example by
extra-dimensional models with
Kaluza-Klein excitations of the electron.   The production and decay process is
illustrated in fig.\ \ref{fig:excited_electron_diagram}.

\begin{figure}[t]
\begin{center}
\includegraphics[scale=0.45,angle=-90]{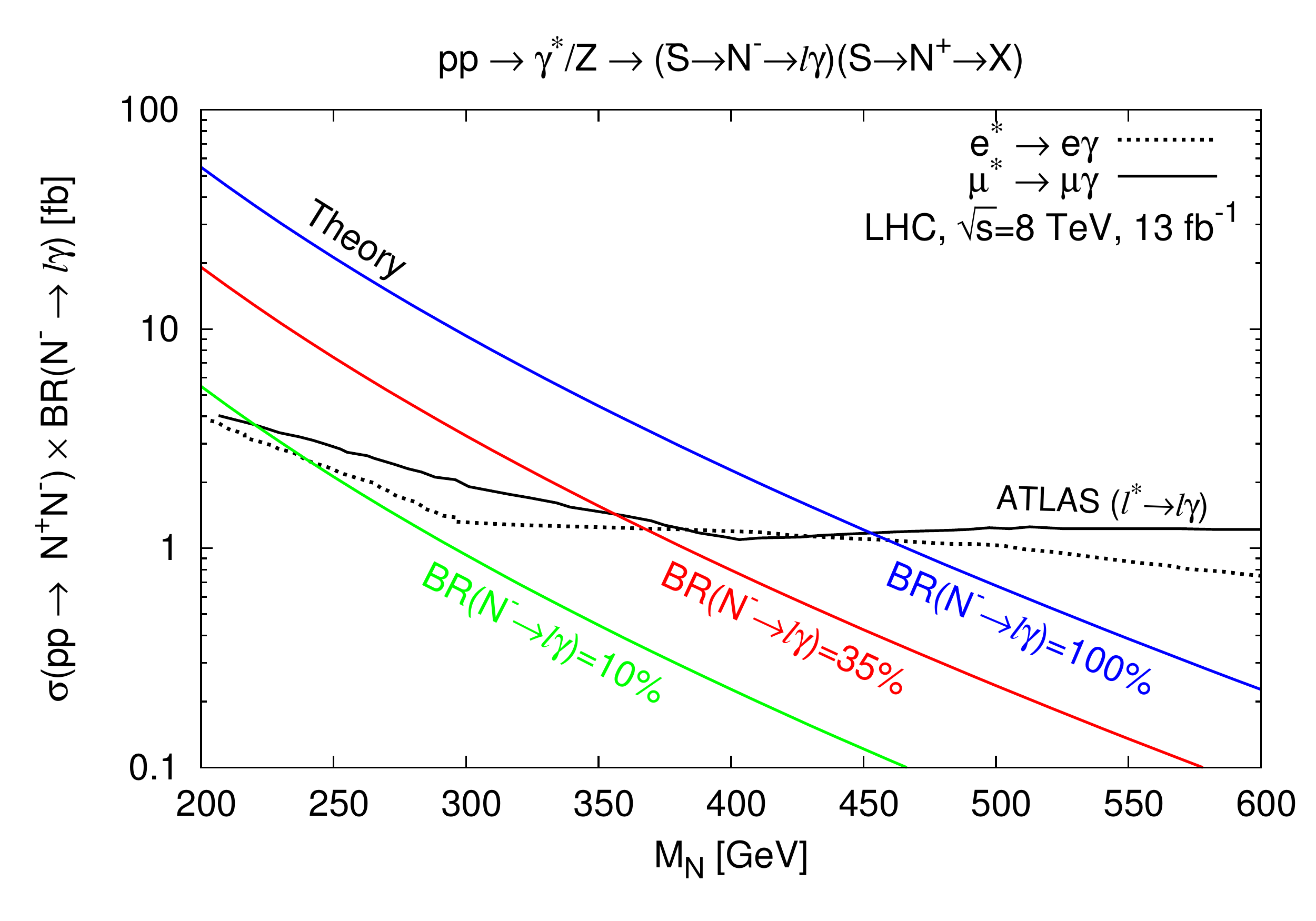}
\caption{Limits resulting from excited electron and muon searches, and
predictions of the MDM model assuming that the branching ratio for
the charged bound state decay
$N^-\to \ell^-\gamma$ is 10\%, 35\% and 100\%, respectively.}
\label{fig:excited_electron_results}
\end{center}
\end{figure}

We have estimated the  production  cross section for $N^+ N^-$ pairs leading to 
$l^\pm\gamma$ final states at the LHC in the MDM model.  In the first approximation,
it is the same as that for producing the unhadronized $\psi\bar\psi$ pair, whose
cross section is 4 times greater than that for producing $SS^*$ at high energies.
Taking into account hadronization is expected to reduce this estimate by a factor 
of $\sim 4$ since there are four possible color-neutral final states involving
$\psi$ ($\psi\psi$, $\psi\bar\psi$, $\psi S$ and $\psi S^*$).
The predictions are shown  
in fig.\ \ref{fig:excited_electron_results} for several representative values of the
branching ratio for $N^-\to \ell^-\gamma$, where $l$ stands for $e$ or $\mu$.   Upper limits on this cross section have been
derived by looking for peaks in the distribution of the invariant mass of the electron
and the photon. ATLAS has searched 
{$\sim 13$ fb$^{-1}$ data at 8 TeV} for this final
state  \cite{ATLAS:2012vja}.  
The resulting constraint for electrons is $\sigma(e^*) \times 
\text{BR}(e^*\to
e\gamma) <(0.6-1)$ fb when $m_{N^\pm}>$ 400 GeV, and for muons $\sigma(\mu^*)
\times \text{BR}(\mu^*\to \mu \gamma) <(0.7-1)$ fb when $m_{N^\pm}>$ 400 GeV.
We will discuss the impact on the MDM model of 130 GeV dark matter in section
\ref{MDMm}.


\section{Two- and four-photon events}
\label{section:4photon}

\begin{figure}[t]
\begin{center}
\includegraphics[scale=0.3]{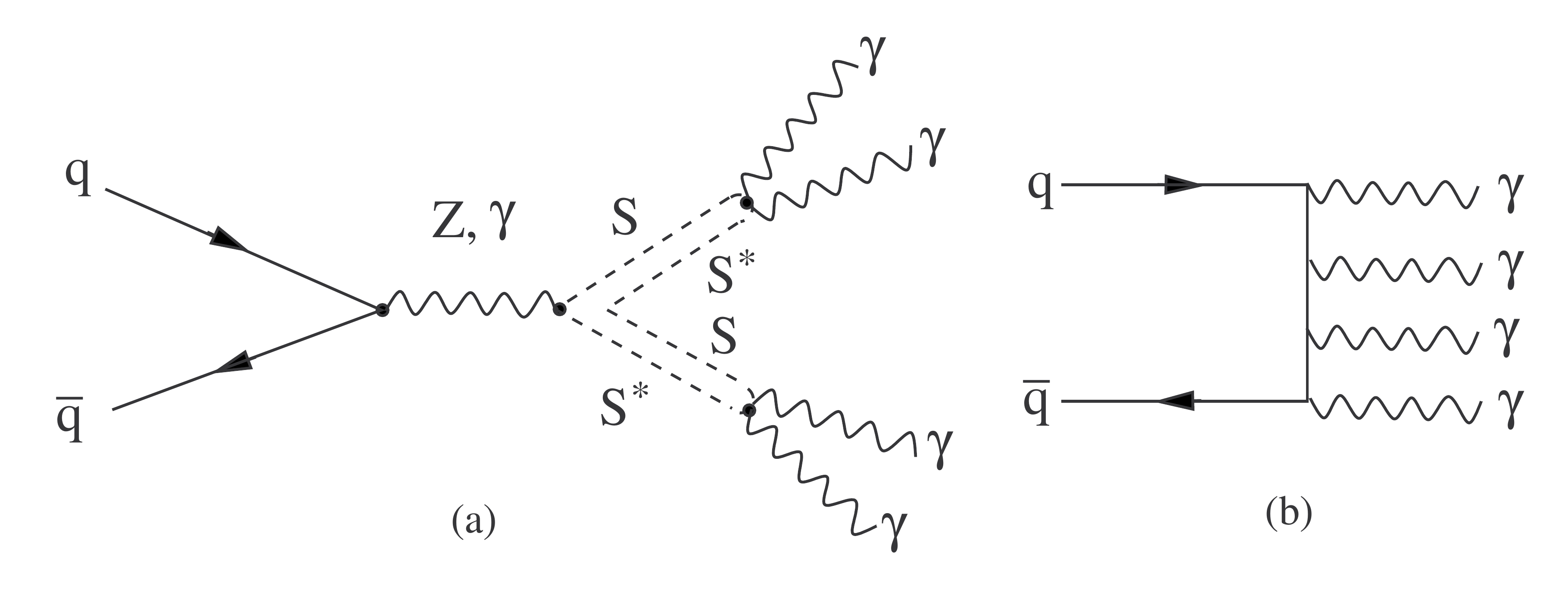}
\caption{(a) (left)  Four-photon event from production of 
$\eta_\S$ pair followed by $\eta_\S\to\gamma\gamma$.  (b)
(right) Higher order standard model background process for four-photon events.}
\label{fig:4photondiagram}
\end{center}
\end{figure}

One of the striking signatures in our models is the 4-photon event
arising from   decays of meson pairs, either $2\eta_\S$ or
$2\eta_\psi$.  Fig.\ \ref{fig:4photondiagram}(a) shows the production
process for $pp\to 2\eta_\S$ followed by $\eta_\S\to 2\gamma$ for each
$\eta_\S$. The branching ratio into photons is significant as long as
there are no lighter hadrons of the SU(N)$_d$ interaction, notably
glueballs, and if decays into dark matter do not dominate.  The first
assumption was previously made in order to forbid 
glueballs as annihilation products of  dark matter, which would
strongly suppress its relic density.  The second one is
model-dependent, as we will discuss in section \ref{viable}.
 The 4-photon
final state is a very clean channel because the primary standard model
background is from analogous QCD processes in which $\pi^0$ and $\eta$
mesons are produced; these photon pairs will not have peaks
in their invariant mass spectra except at very low values, 
and are unlikely to pass the {photon identification criteria at LHC} discussed in more detail later.
There is also a perturbative
contribution to $q\bar q\to 4\gamma$ from fig.\
\ref{fig:4photondiagram}(b), but this is smaller by  $O(\alpha^2)$,
along with other small backgrounds that we discuss below.
The
four-photon signal has not yet been searched for by ATLAS or CMS so we
do not yet obtain any constraints from having two pairs of photons. 
Instead, we make a preliminary study of this channel in section 
\ref{sub:4photon}.  However the four-photon channel will also contribute to existing
searches for single photon pairs and we can use these to set limits,
which we present in section \ref{sub:diphoton}.

\subsection{Diphoton constraints}
\label{sub:diphoton}

Diphotons would be observed from the process shown in 
fig.\ \ref{fig:4photondiagram}(a) since existing searches do not
discriminate against events producing more than two photons.
Here we approximate the $\eta_\S$ meson pair production cross section 
as half that of the elementary $SS^*$ pair in the loop model, 
since there are two ways to hadronize into mesons, whereas in the MDM model
we take production of $\eta_\psi$ pairs to be equal to $1/4$ that
of $\psi\bar\psi$ since there are four possible final states.
We estimate the diphoton production cross section as 
\begin{equation}
\sigma(\gamma\gamma) =  2\,\sigma(pp \to \eta\eta)\, 
\text{BR}(\eta\to \gamma\gamma)
\end{equation}
where $\text{BR}(\eta\to \gamma\gamma)$ is the 
diphoton branching ratio of $\eta$ decays and the factor of 2 accounts
for the two pairs of photons that reconstruct to the right invariant
mass.  In the loop model this may overestimate
the production since we ignore ``baryonic'' final states, $SSS$,
$SST$ and $STT$.

\begin{figure}[t]
\begin{center}
\includegraphics[scale=0.4]{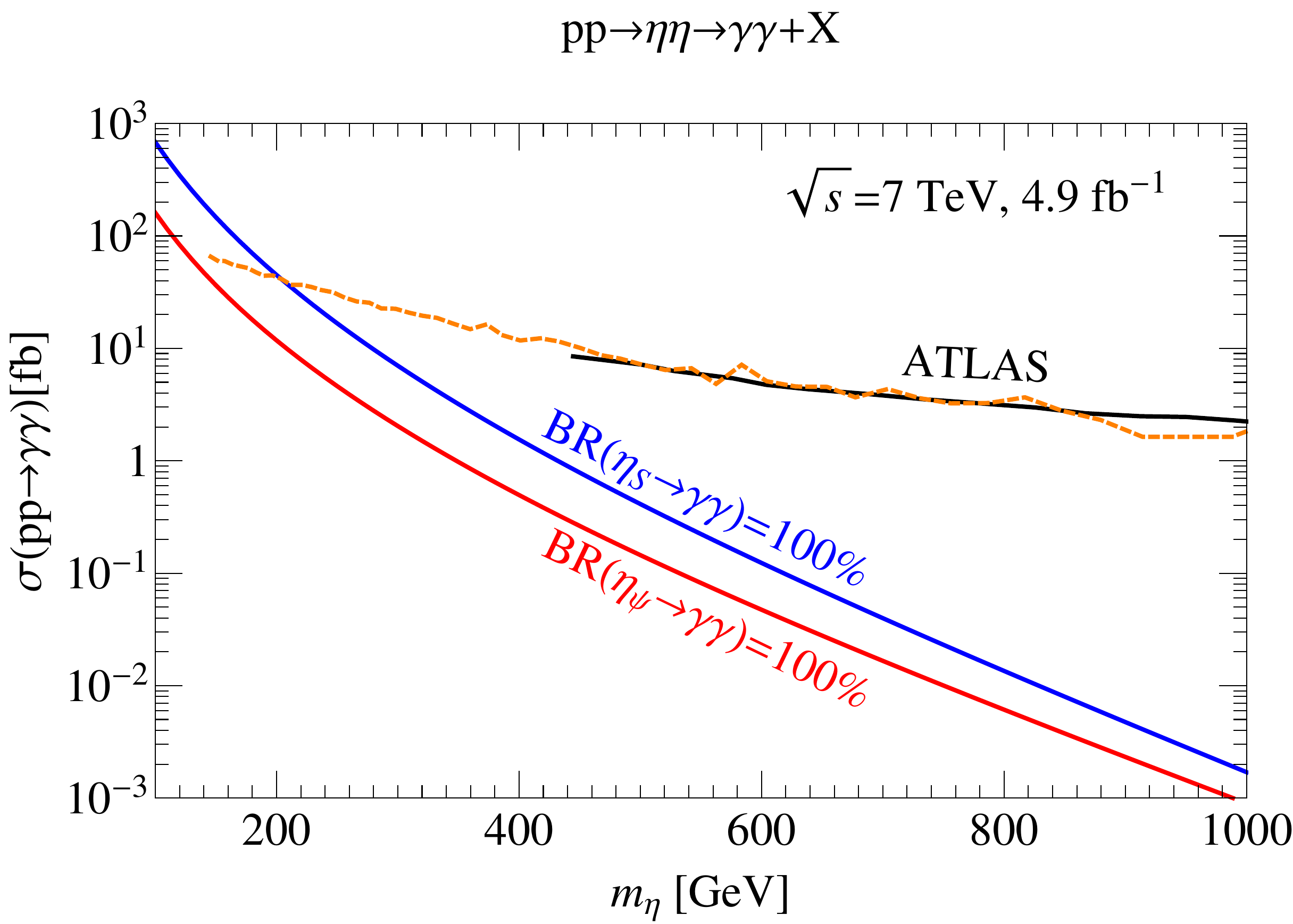}
\caption{{Constraints on the diphoton production cross section from
the ATLAS analysis \cite{ATLAS:2012koa}.  The solid (dashed) ATLAS
curves correspond to the actual (approximated) 95\% C.L.\ limits in
the diphoton channel.  The dashed ATLAS curve which extends the limits
to the low mass range (control region in the  ATLAS analysis) is
estimated using   $\sigma<2\sqrt{N_\text{SM}}/(\mathcal{L}\times
A)$,   where $N_\text{SM}$ is the standard model prediction,
$\mathcal{L}=4.9$ fb$^{-1}$ is the integrated luminosity,  and the
assumed acceptance is chosen to be $A\sim 25\%$ such that the
estimated limits  agree with the actual limits in the high mass
region. The production cross sections from $\eta_S$ and $\eta_\psi$ 
meson decays are also plotted assuming 100\% branching ratio to
diphotons.} }
\label{fig:diphoton}
\end{center}
\end{figure}

Both ATLAS  and CMS  have provided limits on the  diphoton
production cross sections using LHC data at energy $\sqrt{s}=7$ TeV
with integrated luminosity  $\mathcal{L}=4.9$ fb$^{-1}$ (ATLAS)
\cite{ATLAS:2012koa}  and $\mathcal{L}=2.2$ fb$^{-1}$ (CMS)
\cite{Chatrchyan:2011fq}.  Here we apply the ATLAS limits to our
models.  Fig.\ \ref{fig:diphoton} shows the ATLAS constraints on the
diphoton final states in the mass range {$>409$ GeV.  The diphoton
limits in the low mass range 122$-$409 GeV} are estimated based on the
number of  events observed by ATLAS. The theory
predictions of the diphoton signals from $\eta_S$ or $\eta_\psi$
meson decay are also plotted assuming $\text{BR}(\eta \to
\gamma\gamma)=100\%$.  Comparison indicates that meson masses 
{$m_{\eta_\psi} < 120$ GeV and $m_{\eta_\S} < 200$ GeV are
excluded with the integrated luminosity 4.9 fb$^{-1}$. We note that
the diphoton analyses for the SM Higgs boson search near 
120 GeV of ATLAS  \cite{ATLAS:2013oma} and CMS \cite{CMS:ril}
were both  based on the integrated luminosity $\sim$
25 fb$^{-1}$. A dedicated analysis  from these larger data sets would
further improve the limit on $m_{\eta_\psi}$.  }

\subsection{4-photon final state}
\label{sub:4photon}

To simulate the LHC signals of the 4-photon events, we computed the
matrix elements of the  production processes using MadGraph4
\cite{Alwall:2007st}, taking an effective field theory approach to
model the couplings to two photons.  We assumed the 
nonrenormalizable interaction    
\begin{equation} 
	\mathcal{L} = g \eta_\S F^{\mu\nu} F_{\mu\nu}
\end{equation} 
where the coupling strength, $g$, has the
dimensionality of inverse mass.

\begin{figure}[t]
\begin{center}
\centerline{\includegraphics[scale=0.30]{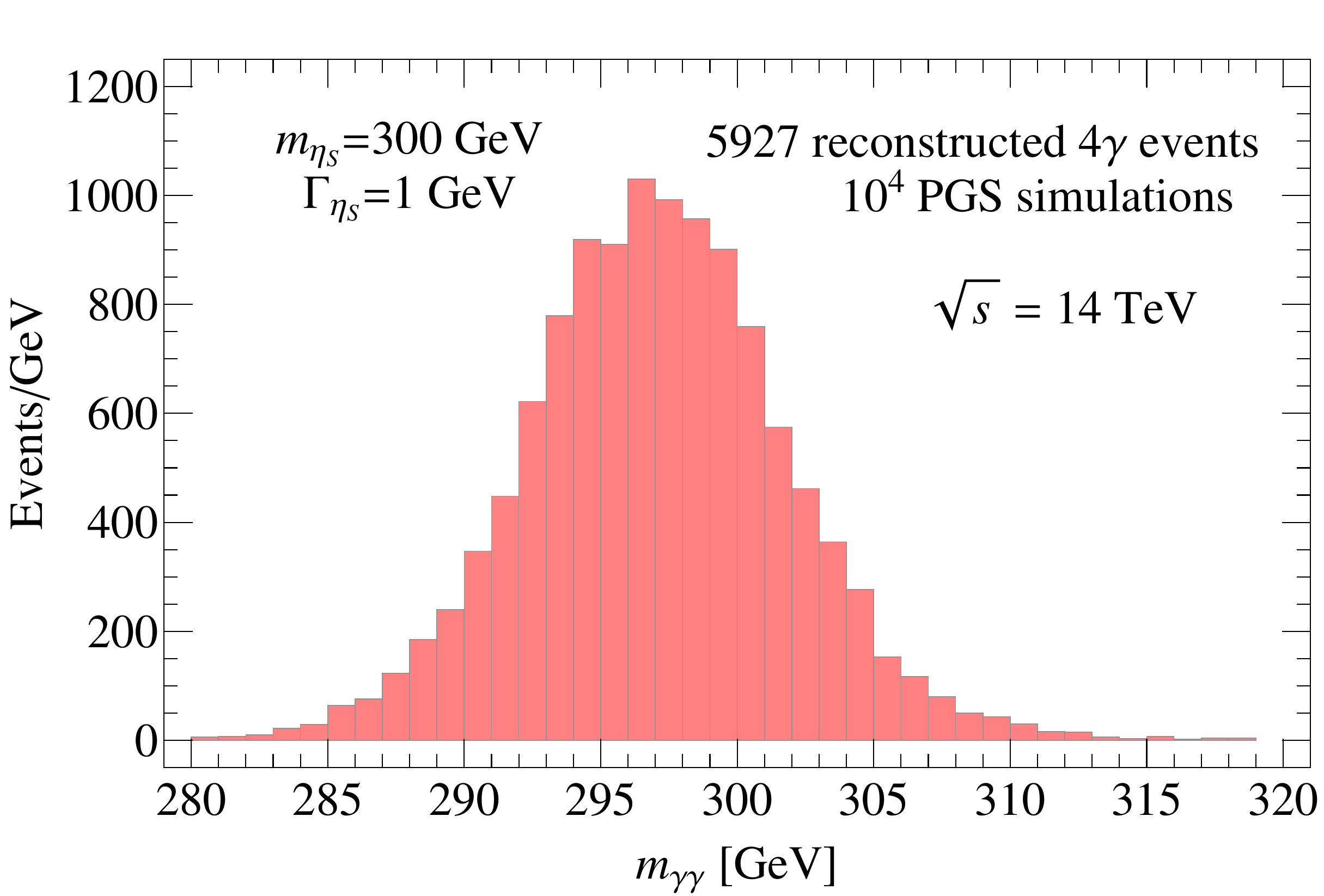}
\includegraphics[scale=0.32]{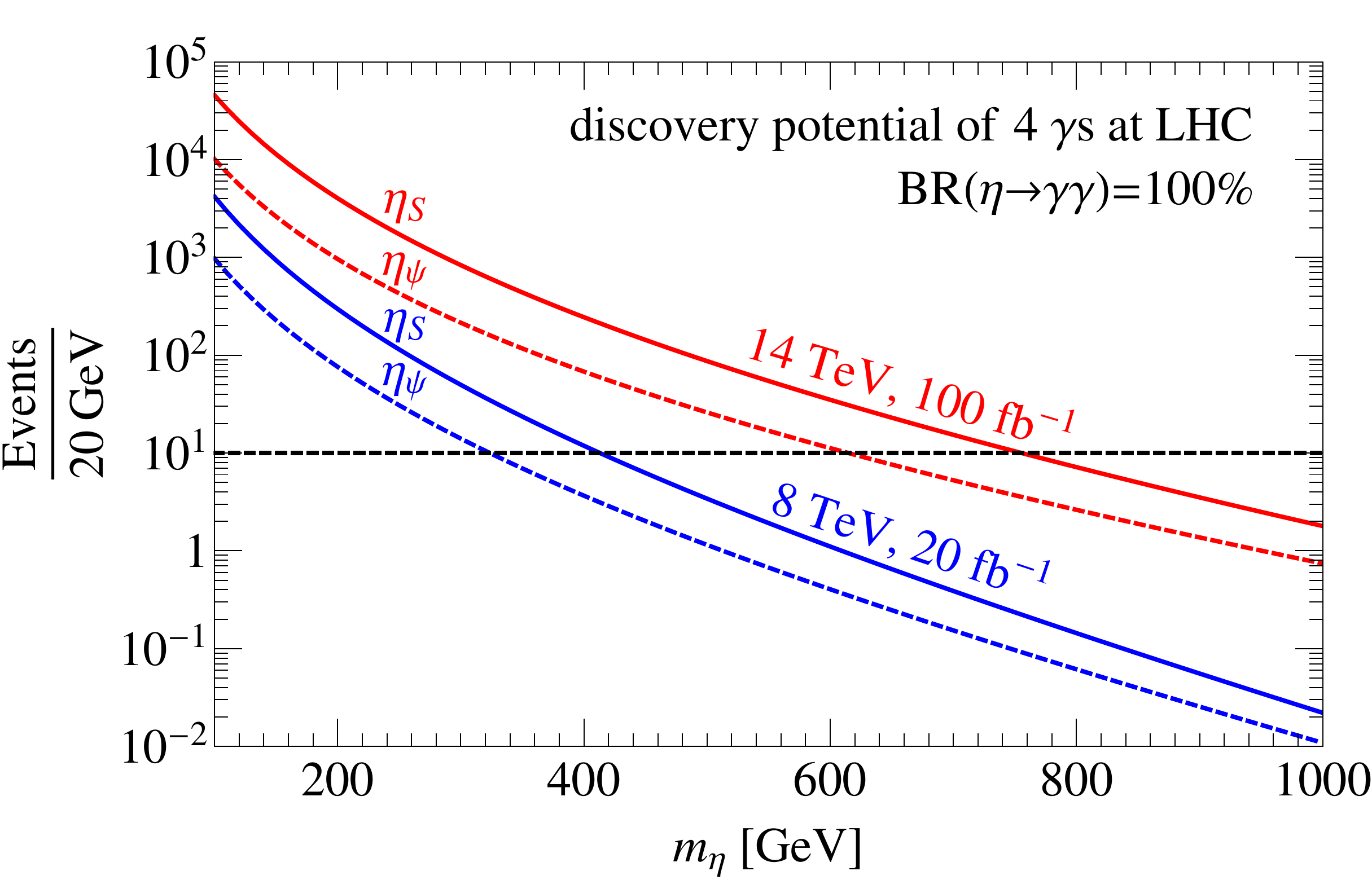}}
\caption{(a) (left) Reconstruction of the 4-photon events from PGS simulation. 
The pairing of photon that results in least mass difference is chosen 
for the invariant mass distribution. 
The shape of the distribution is broadened by the energy resolution of
the detector. 
(b) (right) {Solid (dashed) lines are the} predicted number of 
4-photon events  
{per 20 GeV invariant
mass bin, from $\eta_\S \to\gamma\gamma$ 
($\eta_\psi\to \gamma\gamma$) for 
$\sqrt{s}=14$ TeV, $\int{\cal L}dt = 100$ fb$^{-1}$ (upper pair of curves), 
and for $\sqrt{s}=8$ TeV, $\int{\cal L}dt = 20$ fb$^{-1}$ (lower pair of curves). }
Horizontal line shows the discovery
criterion of 10 events  for reference.
}
\label{fig:4photons}
\end{center}
\end{figure}

We computed the parton level cross sections both for the dark
matter models using the  effective theory approach and for 
the standard model  in MadGraph. The results were then transmitted to PYTHIA
\cite{Sjostrand:2006za} for hadronization and PGS \cite{PGS4} for
collider simulations.  
Following the ATLAS analysis
\cite{ATLAS:2012koa},  we select only the photon candidates with
transverse momentum $p_T>25$ GeV, and the  pseudorapidity in the
ranges $|\eta|<1.37$ or $1.52<|\eta|<2.37$ ({to benefit from  
the high granularity of the first layer in the electromagnetic 
calorimeter 
for discriminating between genuine prompt photons and faked photons within jets}) \cite{Aad:2009wy}. 
An isolation  requirement
on the photon is further imposed: the photon in the  vicinity of  a
jet with $p_T>10$ GeV is vetoed  if the angular distance between the
photon and the jet, $\Delta R = \sqrt{\Delta\eta^2+\Delta\phi^2}$,
is less than 
$0.4$.  After these selection cuts,  only the events that contain 4
photons are considered.  We then search for the correct pairings of
photons, by demanding that they lead to  the smallest difference
between  the two invariant masses out of the 3 possible combinations.
The resulting invariant mass distribution for  the case where
$m_{\eta_S}=300$ GeV and $\Gamma_{\eta_S}=1$ GeV is given in Fig.\
\ref{fig:4photons}(a) as an example of the kind of signal that could
be expected. 
{We then compare the signal events that have both invariant masses 
within a fixed 20 GeV mass bin to the standard model events 
satisfying the same selection criteria to derive the discovery limits 
for this final state. 
The efficiency of this selection for the signal events is 
estimated to be 50\% based on the sample study shown in Fig.\
\ref{fig:4photons}(a).}

In sect.\ \ref{sub:diphoton} we determined the maximum cross section
for diphoton production allowed for the models.  These imply  4-photon
cross sections of {70 fb from $\eta_\psi$ pairs at $m_{\eta_\psi} =
120$ GeV, and 21 fb from $\eta_\S$ pairs at $m_{\eta_\S} = 200$ GeV.
Thus we would predict about 170 and 50} four-photon events respectively in
these two models, for the 4.9 fb$^{-1}$ data set, suggesting that a
dedicated four-photon search could improve the limits.  We have
simulated these events and the background in greater detail in order
to forecast the improvement in constraints, or potential for
discovery, from searching for this signal.  The result is shown in
fig.\ \ref{fig:4photons}(b).  We simulated the SM background shown in fig.\
\ref{fig:4photondiagram}(b) which gives rise to a cross section of only $0.1-0.2$ fb
for $\sqrt{s} = 8-14$ TeV {and $p_T(\text{photon})>10$ GeV in the MadGraph5 \cite{Alwall:2011uj} 
simulation.} Our simulation further shows that only about 2\% of these 
SM 4-photon events can pass the 
selection cuts and the invariant mass requirement, which then contributes 
about only 0.4 events for the case where $\sqrt{s}=14$ TeV and $\mathcal{L}=100$ fb$^{-1}$ 
at the LHC.

To check the SM backgrounds due to contamination from QCD and
electron misidentification,  we further compute the parton level cross
sections in MadGraph for various SM processes  shown in Table.\
\ref{tab:4photon}.  To estimate the faked photons originating from
hadronic jets and isolated electrons,  we use the fake rates (or
background rejection) analyzed by the ATLAS collaboration.  
{The jet backgrounds can be reduced by a factor of $\sim 5000$
using the tight photon selection in ATLAS, with an additional
improvement by a factor of $\sim 1.5$ by adding isolation requirements
\cite{ATLAS:2011kuc}.   The rejection for
quark-initiated and gluon-initiated jets  with
$E_T>20$ GeV are about $1.6\times 10^3$
and $1.4\times 10^4$ respectively \cite{ATLAS:2011kuc}.}   Here we
assume
a conservative  jet rejection based on the quark-initiated jets,
$R\sim 2\times 10^3$, and thus obtain the fake rate  $\sim 5 \times
10^{-4}$ (the inverses of the rejection). For the faked photons due
to  misidentified electrons, we use the measured faked rate
$f_{\gamma\to e} = 0.062$ \cite{Aad:2012tba}.  The effective 4-photon
production cross sections are then computed  taking into account the
faked rates for various SM processes. As shown in Table.\
\ref{tab:4photon},  the 4-photon events at LHC due to contamination
are generally smaller than  the $\gamma \gamma \gamma \gamma$ process
shown in  {fig.\ \ref{fig:4photondiagram}(b)}, and only $\gamma\gamma\gamma j$
and $\gamma \gamma e e$ processes  can yield sizable contributions 
{comparable to} the $\gamma \gamma \gamma \gamma$ process.  
{We further simulated the $\gamma\gamma\gamma j$
and $\gamma \gamma e e$ processes in PYTHIA and PGS for the 
case $\sqrt{s}=14$ TeV and $\mathcal{L}=100$ fb$^{-1}$, but 
no event with invariant mass pair above 100 GeV was found from these 
two SM processes. }

{Since the SM background is thus shown to be very small,} 
we ignore its contribution, and take  $N=10$
events  as the criterion for discovery of 4-photons in these models.  The
Fig.\ \ref{fig:4photons}(b) indicates that the  reach
for discovering  the hidden meson states at LHC is up to $400$ GeV for
$\eta_S$  at $\sqrt{s} = 8$ TeV  with the integrated luminosity
already collected,  and up to $750$ GeV at $\sqrt{s} = 14$
TeV with $100$ fb$^{-1}$ integrated luminosity.  The
corresponding  mass values are somewhat smaller for the 
$\eta_\psi$ due to the smaller electric charge of $\psi$.

\begin{table}[t]
\vspace{5mm}
\begin{tabular}{|c|c|c|c|c|c|c|c|c|c| }
\hline
SM process 
& $\gamma\gamma\gamma\gamma$ 
& $\gamma\gamma\gamma j $ 
& $\gamma\gamma j j $
& $\gamma j j j$
& $j j j j$
& $\gamma\gamma\gamma e \nu$
& $\gamma\gamma e e$
& $\gamma e e e \nu$
& $e e e e$ \\
\hline 
$\sigma_\text{parton}^{\phantom{|}}$  [pb]
& $2\!\cdot\! 10^{-4}$ 
& $2\!\cdot\! 10^{-1}$ 
& $1\!\cdot\! 10^{2}$ 
& $3\!\cdot\! 10^{4}$
& $2\!\cdot\! 10^7$
& $6 \!\cdot\! 10^{-5}$
& $5\!\cdot\! 10^{-2}$
& $9 \!\cdot\! 10^{-4}$
& $7 \!\cdot\! 10^{-3}$ \\
\hline 
$\sigma_{4\gamma}^{\phantom{|}}$  [pb]
& $2\!\cdot\! 10^{-4}$ 
& $1\!\cdot\! 10^{-4}$ 
& $3\!\cdot\! 10^{-5}$ 
& $4\!\cdot\! 10^{-6}$
& $1\!\cdot\! 10^{-6}$
& $4\!\cdot\! 10^{-6}$
& $2\!\cdot\! 10^{-4}$
& $2\!\cdot\! 10^{-7}$
& $1\!\cdot\! 10^{-7}$ \\
\hline
\end{tabular}
\caption{The parton level cross section and  the estimated cross section for the 
4-photon final states for various SM processes at the LHC with $\sqrt{s}=14$ TeV. 
The faked rates $5\times 10^{-4}$ ($0.062$) for photons originating from jets (electrons) are 
assumed here for the estimation of the 4-photon cross sections. The parton level 
cross sections are computed with the transverse momentum cuts:  $p_T\text{(jet)}>20$ GeV, 
$p_T\text{(photon)}>10$ GeV, $p_T\text{(lepton)}>10$ GeV in MadGraph. }
\label{tab:4photon}
\end{table}


\begin{figure}[b]
\begin{center}
\includegraphics[scale=0.35]{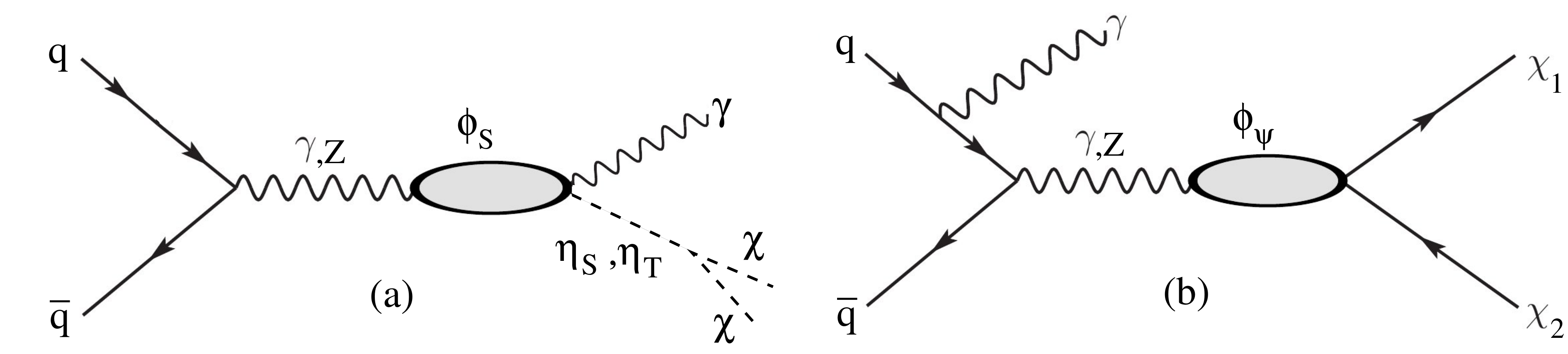}
\caption{Feynman diagrams that generate the monophoton signal
for (a) loop model; (b) MDM model.}
\label{fig:monophoton_diagram}
\end{center}
\end{figure}

\section{Monophoton limits}
\label{section:monophoton}

One way in which constraints on vector mesons  (section
\ref{section:oslepton}) could be evaded is if the branching ratio for
$\phi$ decays into leptons is small due to more dominant decays 
into dark matter or other final states.  In this case, a complementary constraint can be
obtained from searching for monophoton events such as those depicted
in fig.\ \ref{fig:monophoton_diagram}.   In generic dark matter models, the monophoton
arises as initial state radiation (ISR) from the quarks, but for the loop model
that we consider, the vector meson $\phi_\S$ cannot decay only into dark
matter particles, because in this model the DM is scalar and
$\chi\chi$ cannot have the $J^{PC}$ quantum numbers to match those of
$\phi_\S$.  However there is a decay channel
$\phi_\S\to\gamma\eta_\S$ or $\phi_\S\to\gamma\eta_\T$ followed by $\eta_{\S,\T}\to \chi\chi$, as shown in fig.\
\ref{fig:monophoton_diagram}(a), that produces a monophoton and missing energy.
It can naturally dominate over the leptonic decays $\phi_\S\to \ell^+\ell^-$ since it is
lower order in $\alpha$.  It is mediated by the effective operator
\be
g_{\S,\T}\,\frac{e q_\S}{m_\phi}\, \phi_\S^{\mu\nu}F_{\mu\nu}\,\eta_{\S,\T}
\ee 
where we estimate 
{$g_\S = 1.5$} by using the same interaction
to model the charmonium radiative decay processes, $\chi_{c0}\to
\gamma J/\psi$,   $J/\psi \to \gamma \eta_c$, and $\psi(2S) \to \gamma
\chi_{c0}$.  We estimate the ratio 
  $g_\T/g_\S \simeq \alpha_d^2/\pi$, 
  {which encodes the OZI suppression
due to the extra gluon loop for $\eta_\T$ final states},
by comparing the
radiative decays  $J/\psi \to \gamma \eta_c$ and $J/\psi
\to \gamma \eta'$ in QCD in the same model.  {Here $\alpha_d =
g^2_d/4\pi$ is the strength of the SU(N)$_d$ gauge interaction, evaluated at the
scale $m_\eta$ \cite{Korner:1982vg}.}  We take the coupling to run according to the one-loop
beta function,
\be
	\alpha_d(Q^2) = {12\pi\over 32\ln (Q^2/\Lambda_d^2)}
\ee
calculated for SU(3) at scales where $T$ is the only light fundamental
matter.  By comparing to the running of the QCD coupling, we infer
that $\Lambda_d \cong k_d^{1/2}/2$ in the log.

In contrast, monophotons in the MDM model arise from initial
state radiation, but in this case the DM particles produced by decay of the
vector meson $\phi_\psi$ cannot be identical.  This is because the
decay is mediated by the effective
interaction $\phi_\psi^\mu \bar\chi_1\gamma_\mu\chi_2$ that couples 
$\phi_\psi$ 
to the DM ground state $\chi_1$ and the first excited state $\chi_2$
(there are three Majorana DM states in this model, of
which only the least massive one is stable).  The decay to
$\chi_1\chi_1$ is forbidden since the vector current
$\bar\chi_1\gamma_\mu\chi_1$ vanishes identically for Majorana
particles.   The process is depicted in fig.\ \ref{fig:monophoton_diagram}(b). 
Searches for LHC events with a single energetic photon (or jet) 
plus a large missing 
momentum have been frequently used to constrain the 
interaction strength of dark matter with standard
model fermions
\cite{Chatrchyan:2012tea,Chatrchyan:2012me,Aad:2012fw}. 
The signature is a high $p_T$ photon
event accompanied by a significant amount of missing  energy.

In the models we consider, the main parameters controlling the cross
section for monophoton events are the mass $m_\phi$ of the  $\phi_{\S,\psi}$
mesons and their coupling to the photon or $Z$ boson,
as well as the mass of $\eta_{\S,\T}$ in the loop model, or
$\chi_{1,2}$ in the MDM model. In
sect.\ \ref{section:oslepton} we showed that the coupling of the vector
boson to photons was
proportional to the wave function of the origin (for $\phi_\psi$) or
its gradient (for $\phi_\S$).  These in turn depend upon the string
tension $k_d$ of the SU(N)$_d$ gauge interaction through eqs.\
(\ref{psi0eq},\,\ref{gradpsi0eq}).  For low values of $k_d$, such that 
$k_d < m_{\psi,\S}^2$, $k_d$ and $m_\phi$ can be treated as
independent parameters, while for  $k_d \gg m_{\psi,\S}$, we expect
that $m_\phi\sim k_d^{1/2}$, as outlined in appendix \ref{bohr}.  The masses of the final state
particles $\chi_2$ or $\eta_{\S,\T}$  are also determined by $k_d$ when it becomes
large compared to $m_{\psi,\S}$. We expect that $m_{\phi_\S}$ is some multiple
of $m_{\eta_\S}$, roughly independent of $k_d$, while for moderate values of
$k_d$, the $\eta_\T$ can be kept relatively light since $m_\T$ may be
significantly less than $m_\S$.   In the MDM model, $\chi_2$, which is mostly a bound
state of $S\bar\psi$, is expected to have 
$m_{\chi_2}\sim m_{\eta_\psi}$ for large $k_d$.  
In sect.\ \ref{MDMm} we will show that it is possible to
make $m_{\chi_1}$ parametrically smaller than $k_d^{1/2}$ even when
$k_d$ is large.

\begin{figure}[t]
\begin{center}
\includegraphics[scale=0.3]{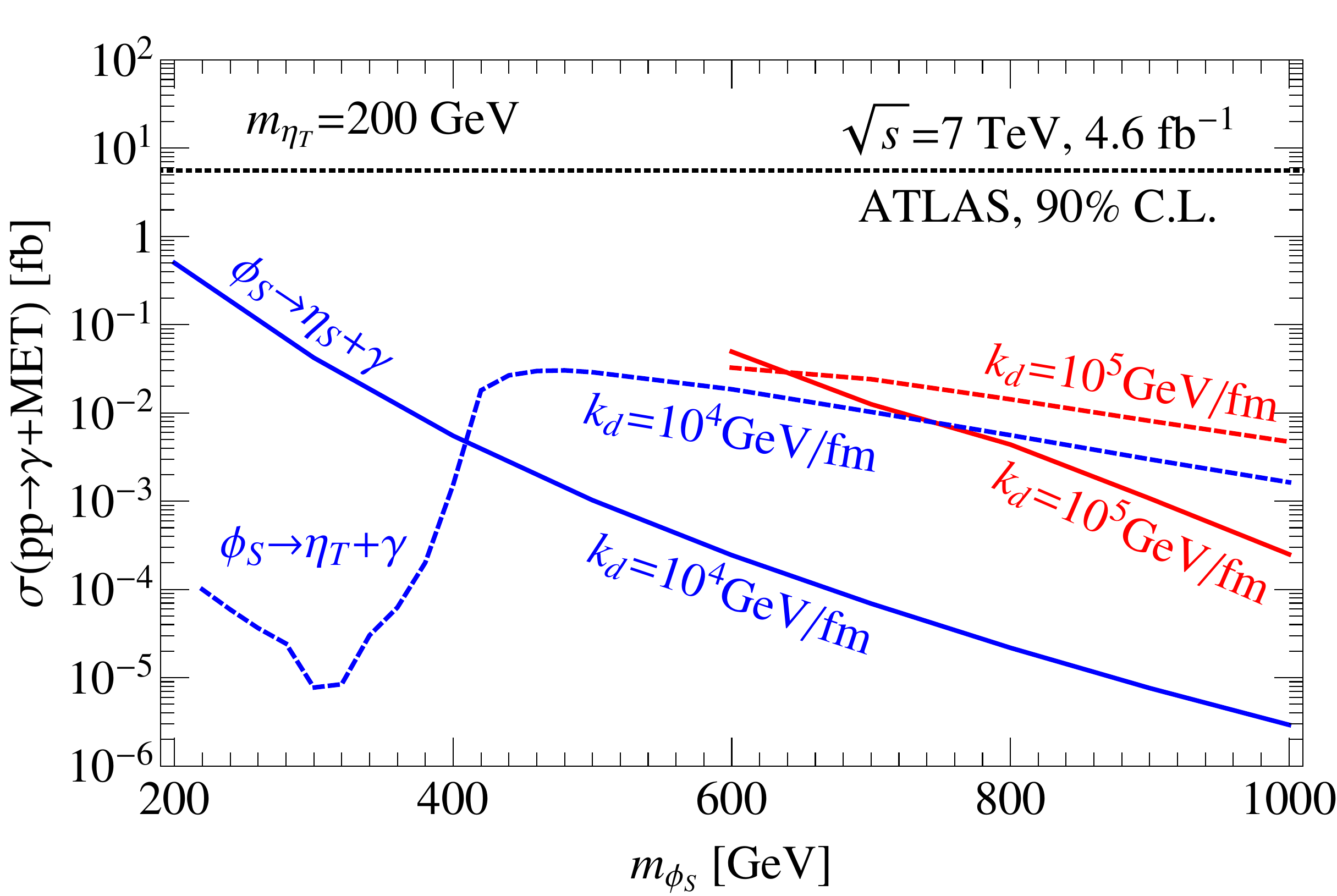}
\includegraphics[scale=0.3]{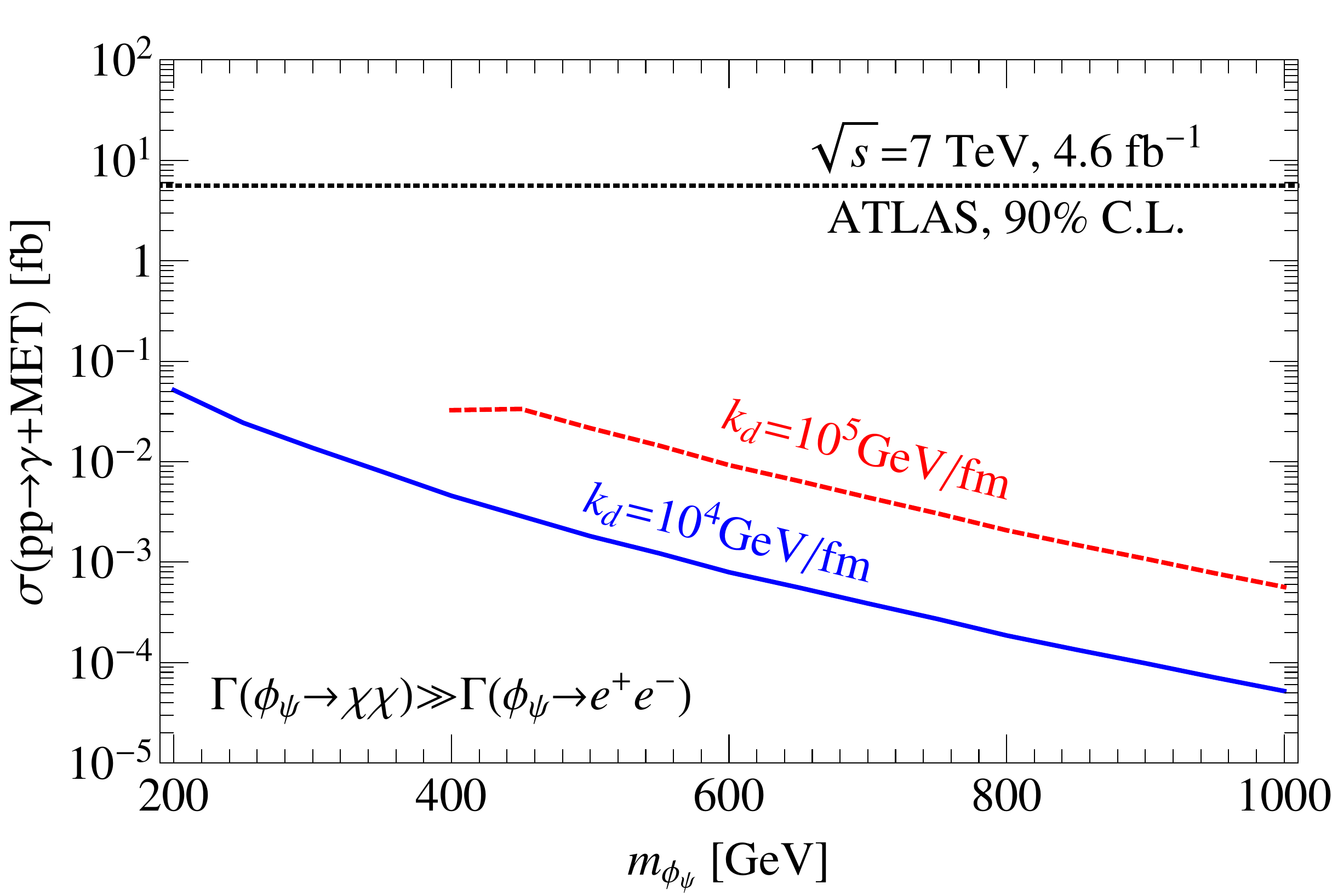}
\caption{
{(a) (left) Upper line (dotted): ATLAS constraint on the cross section for 
monophoton production.  Dashed curves: predictions of loop model for
$\phi_\S\to \eta_\T+\gamma$ at the values $k_d=10^4, 10^5$ GeV/fm of the SU(N)$_d$ string tension.  
Solid curves: similar to dashed but for $\phi_\S\to \eta_\S+\gamma$.
(b) (right) analogous monophoton constraints from ISR in the MDM model, at $k_d=10^4, 10^5$ GeV/fm.}
}
\label{fig:monophoton_constraint}
\end{center}
\end{figure}

To determine the LHC constraints from monophotons, we first
computed the parton level cross sections for the processes shown in fig.\
\ref{fig:monophoton_diagram} using MadGraph. 
The detector acceptance and efficiency were determined using
 PYTHIA and PGS simulations. 
Following the ATLAS analysis \cite{Aad:2012fw}, a minimum photon $p_T$ 
of 80 GeV was required for the event simulations with MadGraph. 
For the  
events simulated with PYTHIA and PGS, we imposed the following sets of selection cuts: 
(1) events were required to have $E_T^\text{miss}>150$ GeV; 
(2) a photon was required with $p_T>150$ GeV and $|\eta|<1.37$ or $1.52<|\eta|<2.37$; 
(3) events with more than one jet with $p_T>30$ GeV and $|\eta|<4.5$ 
were rejected; 
(4) events with identified electrons (muons) with $p_T>20$ GeV 
and $|\eta|<2.47$
($p_T>10$ GeV and $|\eta|<2.4$) were rejected;  
(5) The angular separations between the photon, the missing transverse
energy, and the jet, $\Delta\phi(\gamma, E_T^\text{miss})$, 
$\Delta R (\gamma, \text{jet})$, 
$\Delta\phi(\text{jet},E_T^\text{miss})$, were all required to be larger than 0.4. 
For the MDM model with the vector meson mass in the range 200$-$1000 GeV, 
the detector efficiency for our simulations is found to be 
$A\times \epsilon \sim$(14\%$-${27\%}), which is in the same range as the detector efficiencies  
for {the} various DM effective operator interactions in the ATLAS analysis \cite{Aad:2012fw}.

The LHC production cross section for monophoton final states taking 
into account the detector effects was then computed for the case where 
the vector meson decays into hidden sector fermion pairs (MDM model), 
or into a single photon plus two dark matter particles (loop model). 
{In the loop model, we computed the branching ratios of the $\phi_S$ decaying  
into $\gamma\eta_\S$ and $\gamma\eta_T$, and found that these
dominate over the decays into
SM final states. In the MDM model, we assume that the invisible decays
of the $\phi_\S$ dominate.}
As shown in fig.\ 
\ref{fig:monophoton_constraint}, we find that the monophoton constraint is 
more stringent for the loop model than for the MDM model.  
This is partly because the
cross section for the latter process is down by $O(\alpha)$, since the photon
comes from the initial state, whereas it arises from the decay of $\phi_\S$
in the former. 
{This statement is true however only if the final state photon
is energetic enough to pass the imposed cut, $E_\gamma > 150$ GeV.  
Based upon the ``Bohr model'' predictions of appendix \ref{bohr}, we find that
this is not the case for $\phi_\S\to\eta_\S+\gamma$,
whereas it is true for
$\phi_\S\to\eta_\T+\gamma$ at sufficiently high $m_{\phi_\S}\gtrsim 350$
GeV, if we assume that $m_{\eta_\T}$ is fixed to be $200$ GeV.  This explains
the qualitative behavior of the
$\phi_\S\to\eta_\T+\gamma$ curve for $k_d=10^4$ GeV/fm in fig.\ 
\ref{fig:monophoton_constraint}(a). 
None of these predicted cross
sections conflicts with the ATLAS upper limit.}

The monophoton signals arising  from the MDM model with the string tension in
the range $10^4-10^5$ GeV/fm  are also significantly below  the ATLAS 90\% C.L.\
limit, 5.6 fb, as shown  in fig.\ref{fig:monophoton_constraint}(b).  For higher
$k_d$, for example $10^6$ GeV, the mass of $\phi_\S$ exceeds 1 TeV and so the
resulting prediction would not appear in the range of masses plotted  (and
moreover would still be below the upper limit).   Thus the MDM model is not
constrained by the current monophoton searches.


\section{Viability of 130 GeV (and other) DM models}
\label{viable}

In this section we combine the preceding LHC constraints with the
requirements from sect.\ \ref{section:models} of the Fermi 130 GeV line to determine what regions of
parameter space of the models under consideration are compatible
with all the data.  After considering the three classes of models,
we sketch how the most relevant of these results would generalize to
potential dark matter candidates of other masses, resulting in
gamma ray lines at different energies.

\subsection{Loop model}
\label{loopm}

In order to get a large enough cross section for
$\chi\chi\to\gamma\gamma$, eq.\ (\ref{svl}) implies that $m_S$ must
not be too much larger than $m_\chi$, so that the loop is resonantly
enhanced; otherwise nonperturbatively large values of the coupling
$\lambda_{\S\chi}$ are required.  To make the r.h.s.\ of (\ref{svl})
equal to 1 for the fiducial values of parameters given there, one
finds that $m_\S = 147$ GeV.  Smaller values of $m_S > 130$ GeV
would allow for $\lambda_{\S\chi}<3$, with a minimum value of 
$\lambda_{\S\chi}=1$ when $m_\S$ is just above 130 GeV.  Thus
the range of allowed values is limited to $m_\S\sim 130-147$ GeV.

The constraint from same-sign dileptons, fig.\ \ref{fig:ssdilepton},
can be evaded if
the branching ratio of $\eta_{\S\T}$ into $ee$ or $\mu\mu$ is
sufficiently small, or if $m_{\eta_{ST}} > 200$ GeV in the case of
100\% branching to $\tau\tau$.  The models we
consider do not specify the relative couplings of $\eta_{\S\T}$ into
different flavors of leptons, but if we adopt some version of 
minimal flavor violation, it would be natural to expect that the
partial widths for $ee$, $\mu\mu$ and $\tau\tau$ final states
are in the ratio $m_e^2:m_\mu^2:m_\tau^2$.  Thus we require that
$m_{\eta_{ST}} > 200$ GeV.  Since $m_\S>130$ GeV, this will be 
satisfied if $m_\T > 70$ GeV, and perhaps for smaller values, depending upon how
much of $m_{\eta_{ST}}$ is due to the gluons of SU(N)$_d$.

\begin{figure}[t]
\begin{center}
\includegraphics[scale=0.4]{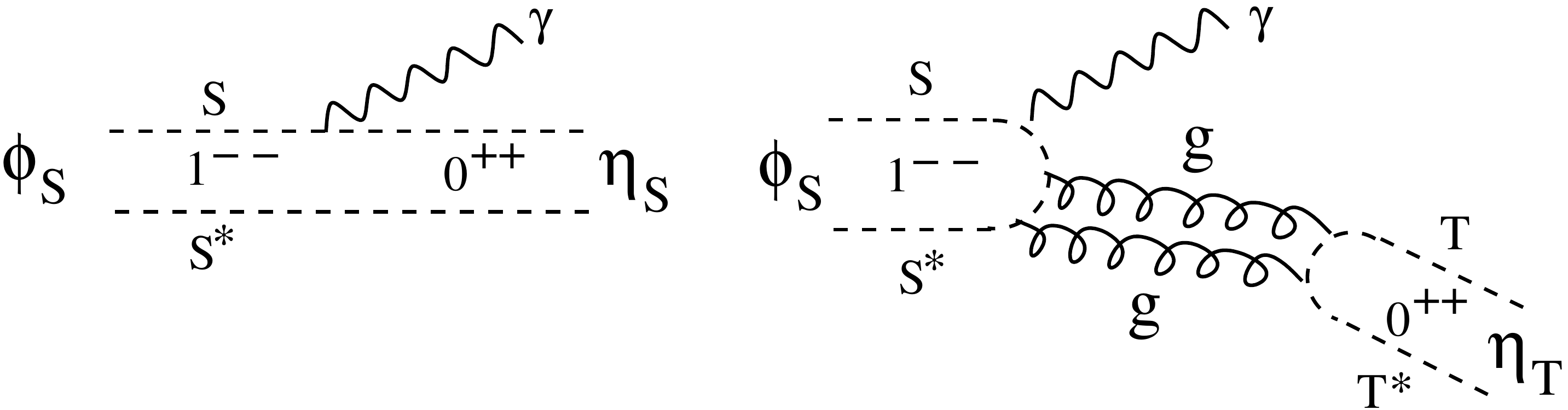}
\caption{Dominant decays of $\phi_S$ vector meson, which reduce the
branching ratio for leptonic final states.  The second process, where
$g$ stands for the SU(N)$_d$ gluons, exists only
in the loop model, while the first is present in both models.}
\label{fig:phiS-decay}
\end{center}
\end{figure}

Direct production of the vector meson $\phi_\S$ followed by its decay
into leptons is potentially constraining, but this depends
upon the branching ratio of $\phi_\S$ into leptons (and into quarks,
because of the production part of the amplitude).   In fact, there are
more dominant decay channels $\phi_\S(1P) \to \eta_\S(1S)+\gamma$ and 
$\phi_\S(1P) \to \eta_\T(1S) + \gamma$ shown in
fig.\ \ref{fig:phiS-decay}.  These are lower order in the
electromagnetic coupling than $\phi_\S\to \ell^+\ell^-$ and are therefore
expected to have a larger branching ratio by of order $1/\alpha$.
The diagram 
with gluons is not suppressed (by the Okubo-Zweig-Iizuka (OZI) rule) in the case where the constituent masses
are at the confinement scale.  Taking $0.03$ as a representative 
value of the leptonic branching ratio,\footnote{\label{brfn}
This number is
motivated by the example $\Gamma(\phi\to
e^+e^-)/\Gamma(\phi\to \eta\gamma) = 3\times 10^{-2}$ from QCD}
 we infer from fig.\ \ref{fig:dilepton} the lower limit
{$m_{\phi_\S}\gtrsim 310$ GeV,} 
which requires $\phi_\S(1P)$ to get 
{$\gtrsim 5-15\%$} 
of its mass from the strong interactions as
opposed to the constituent masses $m_\S$.   The meson mass is given
by $m_\phi\cong 2 m_\S \lesssim 300$ GeV only if $m_\S^2\gg k_d$
(the string tension of SU(N)$_d$), but we have already argued that
this regime is not allowed since otherwise dark matter would
annihilate too strongly into glueballs of SU(N)$_d$.  In the opposite
limit $m_\S^2\ll k_d$, the meson mass scales as $\sqrt{k_d}$
(appendix \ref{bohr}).  Thus only a modest hierarchy in $k_d/m_\S^2$
is needed to satisfy the dilepton constraint.  If $k_d/m_\S^2$ is
not too large, this constraint would be close to saturation, hence
leading to discoverable new physics.

The diphoton searches provide a weaker limit $m_{\eta_\S} >
220$ GeV (fig.\ \ref{fig:diphoton}), which is automatically satisfied
due to the constraint $m_\S > 130$ GeV.  It might be possible
to improve this limit by doing a dedicated search for 4-photon events as we
have suggested in sect.\ \ref{sub:4photon}.  However, the strength of
these constraints depends upon the branching ratio for
$\eta_\S\to\gamma\gamma$.  In this model, the invisible decays to 
dark matter, $\eta_S\to\chi\chi$, are guaranteed to dominate since
\be
	{\Gamma(\eta_\S\to\gamma\gamma)\over 
	\Gamma(\eta_S\to\chi\chi)} \cong
{\sigma(SS\to\gamma\gamma)\over
	\sigma(SS\to\chi\chi)} \sim {(2e)^4\over\lambda_{\S\chi}^2}
	\sim {0.13\over 9}
\ee
Recall that the coupling $\lambda_{\S\chi}$ had to be rather large in
order to get the observed galactic gamma ray line strength.

\subsection{MDM Model}
\label{MDMm}

From eq.\ (\ref{svm}) one finds that the composite magnetic dark 
matter model requires the mass of the fermionic constituent to be
 $m_\psi\sim 70$ GeV 
in order to get a large enough magnetic moment to match the galactic
observations.  While this sounds dangerously low, the model
has the interesting possibility of
allowing for the confinement scale $\Lambda_d$ to be parametrically
higher than the DM ground state $m_{\chi_1}=130$ GeV, because $\chi_1$ is a mixed state of an
elementary Majorana fermion $\chi$ (neutral under the SU(2)$_d$) and the
Dirac bound state $\eta = S\psi$.  The mixing is provided by the
Yukawa interaction $y\bar\chi S \psi$ in the fundamental Lagrangian,
leading to an off-diagonal mass term $m_y \bar\chi\eta$ in the
low-energy effective theory, with $m_y\sim y\Lambda_d$.  Considering
for illustration the case where $m_\eta = m_\chi$ ($m_\chi$ being
the mass of
the original unmixed $\chi$ field), corresponding to maximal mixing,
one has
\be
	m_{\chi_1} = m_\eta - \sqrt{2} m_y
\label{dmmass}
\ee
(see eq.\ (C1) of ref.\ \cite{Cline:2012bz}).
Thus if $m_\eta\sim \Lambda_d$ happens to be close to $\sqrt{2} m_y$,
there can be a moderate hierarchy in $\Lambda_d/m_\psi$ without 
excessive fine-tuning.  This freedom can help the model to satisfy
the LHC constraints, since then $\psi$ only appears in bound states 
that are significantly heavier than its rather low bare mass. 

First we consider the direct production of the vector meson $\phi_\psi$
followed by decays into leptons.
In addition to the
fermionic decay channels, there is a lower-order decay process $\phi_\psi(1S)
\to \eta_\psi(1S)+\gamma$, similar to fig.\  \ref{fig:phiS-decay},
except now the final hadron has $0^{-+}$ quantum numbers.  Therefore
the $\eta_\psi(1S)$ and $\gamma$ must be in an $l=1$ state to conserve
parity.  This could in principle suppress the rate, since the
effective interaction requires an extra derivative,  but if the mass
splitting between $\eta_\psi$ and $\phi_\psi$ is of order the
confinement scale, as would be the case assuming the moderate hierarchy
between $\Lambda_d$ and $m_\psi$ mentioned above, there is no
significant
suppression.   The analogous situation is observed in QCD for the
decays of the $\bar s s$  vector meson $\phi$; see footnote
\ref{brfn}.   Taking $0.03$
as the leptonic branching ratio gives the constraint 
{$m_{\phi_\psi} \gtrsim 250$ GeV.}  
In view of the preceding discussion, by taking
$\Lambda_d\sim 300$ GeV, it would not
be difficult to satisfy this constraint while keeping the lightest
DM state at 130 GeV.

Next we turn to the limits from excited lepton searches.
As  mentioned in section \ref{section:excited}, there is a transition 
magnetic moment between the charged bound states $N^-=S^*\psi$ and the
right-handed leptons in the mass eigenstate basis. 
The mass mixing
arises from the Yukawa interaction 
$\yl\epsilon_{ab}S^*_a\bar \ell_\R\psi_b$  which in the low-energy theory
can be represented by a mass term $\yl \fN \bar \ell_\R N^-_\L$, 
where
$\langle 0|\epsilon_{ab}S^*_a\psi_b|N^-\rangle = \fN$.   For
simplicity consider the case where only a single generation of leptons
mixes significantly with $N^-$.  For small $\yl$, the mixing angle is 
$\theta\cong \yl \fN / m_\N\sim \yl,$ since by analogy with pions
we expect $\fN/m_\N\sim 1$.  The transition magnetic moment is
given by $\mu = \theta(\mu_\ell-\mu_\N)\cong \theta\mu_\ell$, since we
expect that $\mu_\ell \gg \mu_\N$ due to the fact that $m_\ell\ll 
m_\N$.
The partial width for $N^-\to \ell^-\gamma$ is thus of order  
\[	\Gamma(N^-\to \ell^-\gamma) \sim {y_\ell^2\mu_\ell^2\over 8\pi} m_\N^3 \]
Interestingly, if we make the assumption of minimal flavor violation
so that 
$y_\ell\sim m_\ell$, the flavor dependence cancels out since
$\mu_\ell\sim e/m_\ell$.  In this case the branching ratio is roughly equal to
all lepton flavors.  The search for excited electron decays (fig.\
\ref{fig:excited_electron_diagram}) 
then gives the limit $m_N > 367$ GeV, which is somewhat stronger than 
the dilepton constraint.  Nevertheless it can be accommodated by
taking the confinement scale to be of this order.

\begin{figure}[t]
\begin{center}
\includegraphics[scale=0.4]{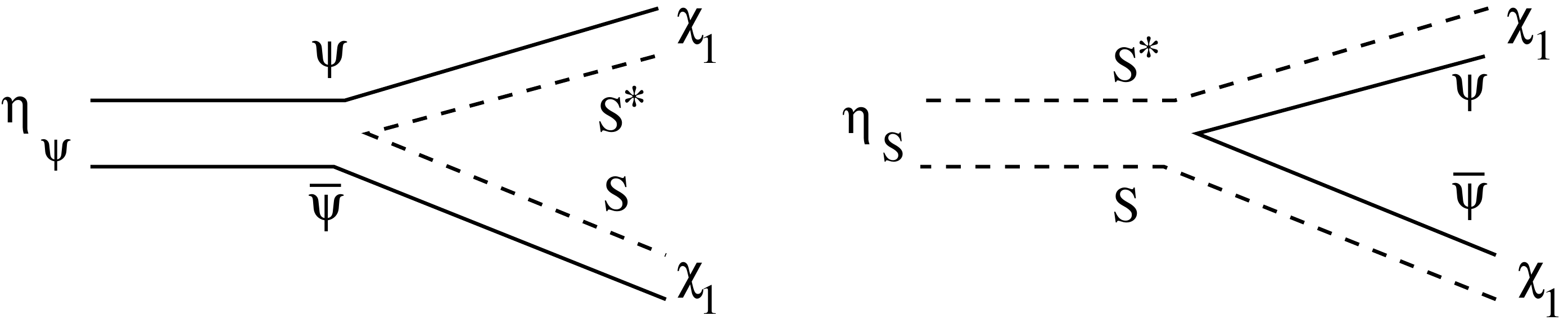}
\caption{Dominant decays of $\eta_\psi$ and $\eta_\S$ mesons into dark
matter in the MDM model, which reduce the
branching ratio for diphoton final states. The dark matter ground
state $\chi_1$
is actually an admixture of the $\psi S^*$ or $\bar\psi S$ bound state and a
neutral Majorana fermion $\chi$.}
\label{fig:eta-psi-S-decay}
\end{center}
\end{figure}

The diphoton constraint from fig.\ \ref{fig:diphoton}, 
$m_{\eta_\psi}>
140$ GeV,  is relatively weak, but as in the case of the loop model,
it might be leveraged into a more sensitive test if 4-photon events were
analyzed in the data.  Again, this depends on the branching ratio for
$\eta_\psi\to\gamma\gamma$.  Like in the loop model, there exist more
dominant decays of $\eta_\psi$ (or $\eta_\S$) into two dark matter
particles.  At the parton level, this is mediated by the diagrams in
fig.\ \ref{fig:eta-psi-S-decay}, which are strong decays and not
suppressed by any small couplings, but only by the mixing angle
between the $\bar\psi S$ bound state and the $\chi_1$ mass eigenstate
(recall that $\chi_1$ is an admixture of $\bar\psi S$ and a neutral
Majorana fermion $\chi$).  This mixing angle is typically large,
$\pi/4$ in the maximal mixing case exemplified in eq.\
(\ref{dmmass}).  Thus we can expect the branching ratios for
$\eta_{\psi,\S}\to\gamma\gamma$ to be suppressed by $O(e/2)^4\sim
10^{-3}$, and the monophoton signal becomes potentially more important
for constraining the invisible decays of $\eta_{\psi,\S}$ to dark
matter.  However we found in fig.\ \ref{fig:monophoton_constraint}(b)
that in fact this constraint is also weak and does not yet restrict the model.

\subsection{$s$-channel regime}

As we discussed in sect.\ \ref{s-chan}, 
both of the models considered here encompass an alternate picture
for explaining
the 130 GeV line, when the $\eta_\S$ or $\eta_\psi$ meson has mass
close to {260} GeV, due to the resonant annihilation 
$\chi\chi \to \eta\to 2\gamma$.  In the case of the loop model, this
is just another way of viewing the loop when it is resonantly
enhanced, giving results that are compatible with the perturbative 
analysis.  In the magnetic model however it is a qualitatively
different mechanism, that does not  rely upon the transition
magnetic moment of the dark matter states.  One can consider larger
values of $m_\psi$ that mildly suppress the magnetic moment, and allow
the confinement scale to be smaller so that $\eta_\psi$ gets most of
its mass from the $\psi$ constituents.  However $m_\psi$ cannot exceed
130 GeV, so quantitatively this regime is not far separated from the
one previously considered.  In any case, the LHC constraints are not
more difficult to satisfy in this scenario, and it comes with the
added prediction that the diphotons produced in 4-photon events 
should have invariant masses near {260} GeV.  This is in the range we
estimated to be reachable during the next LHC run.

\subsection{Beyond 130 GeV}
\label{beyond}

\begin{figure}[t]
\begin{center}
\includegraphics[scale=0.30]{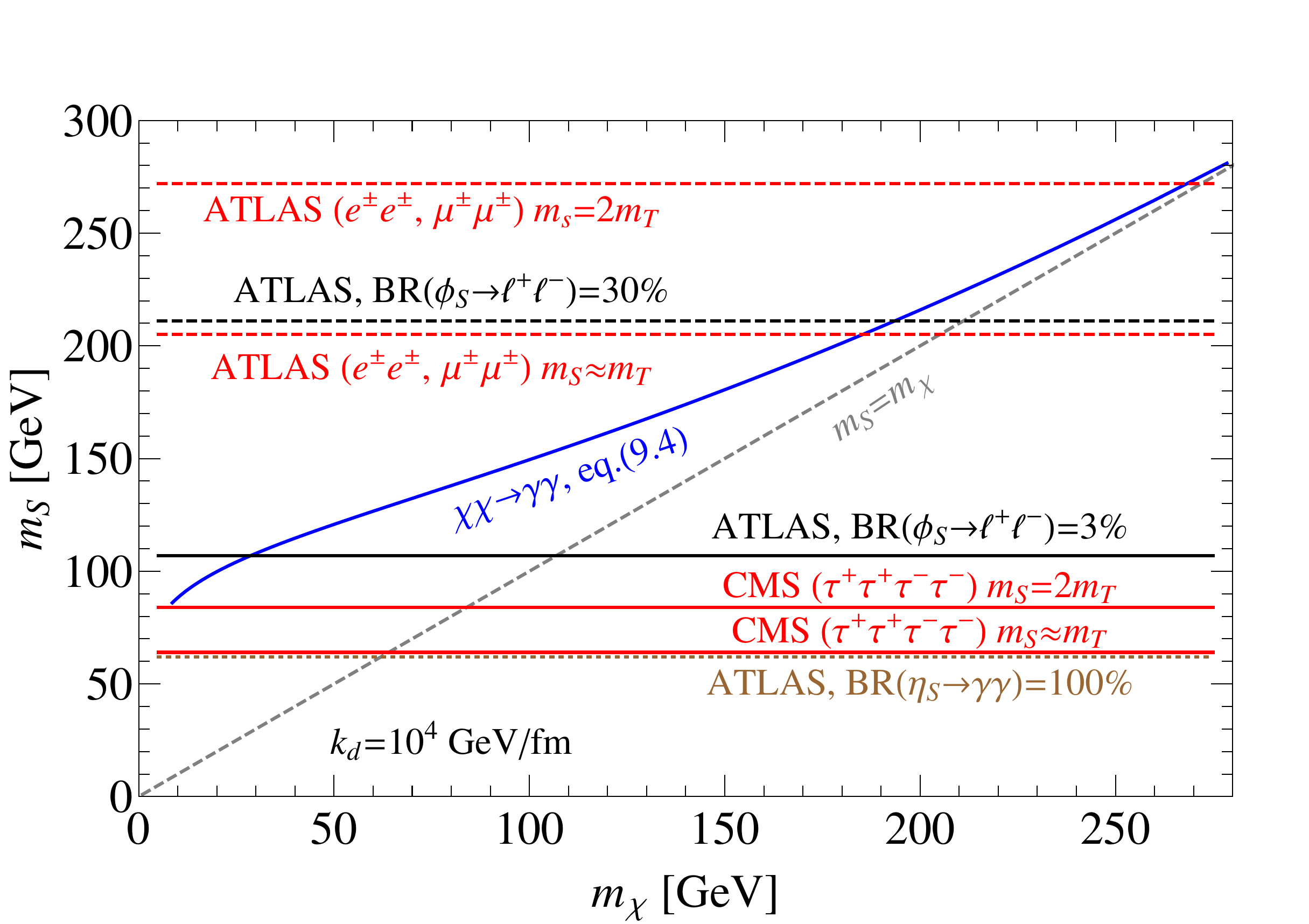}
\includegraphics[scale=0.30]{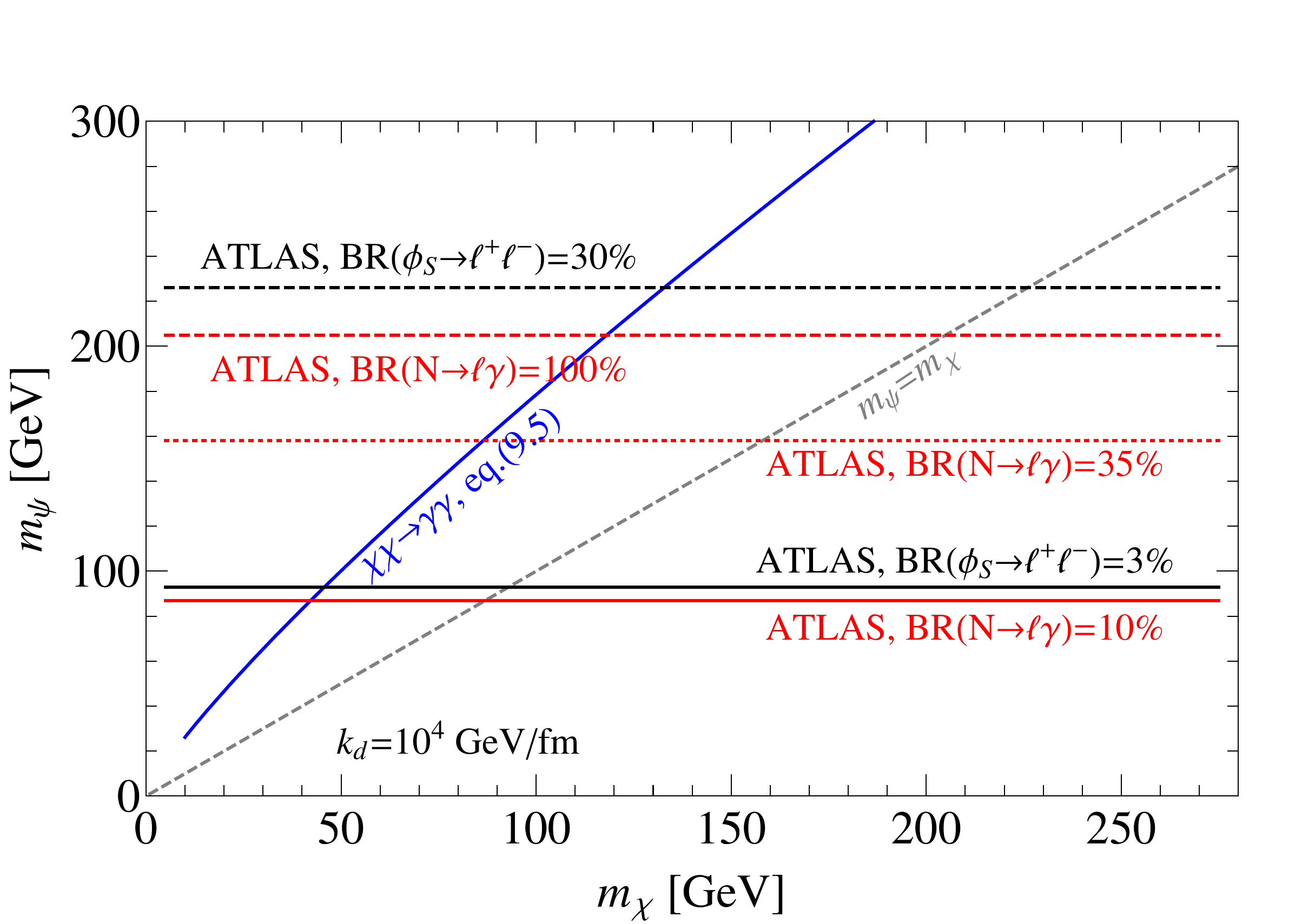}
\includegraphics[scale=0.30]{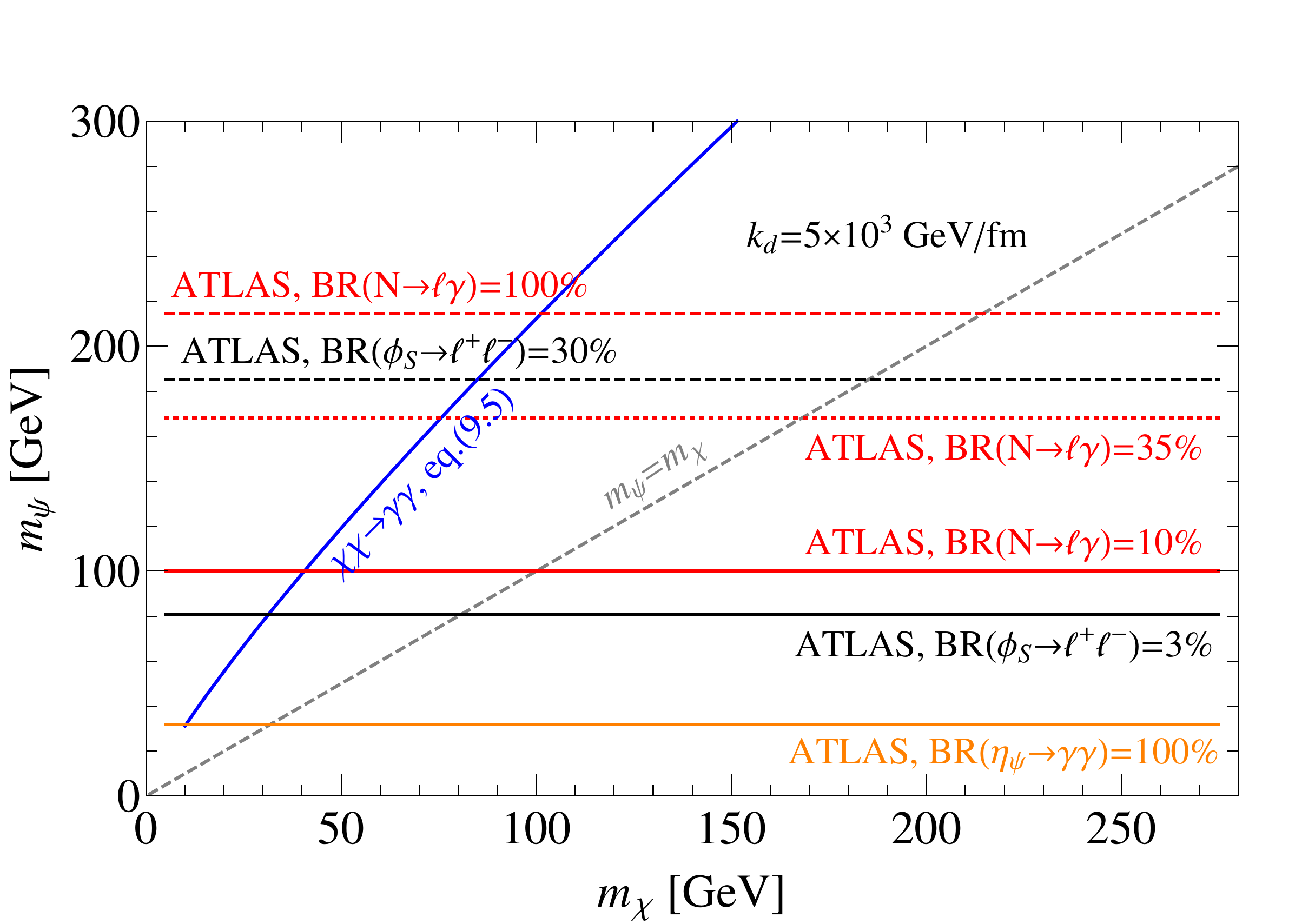}
\includegraphics[scale=0.30]{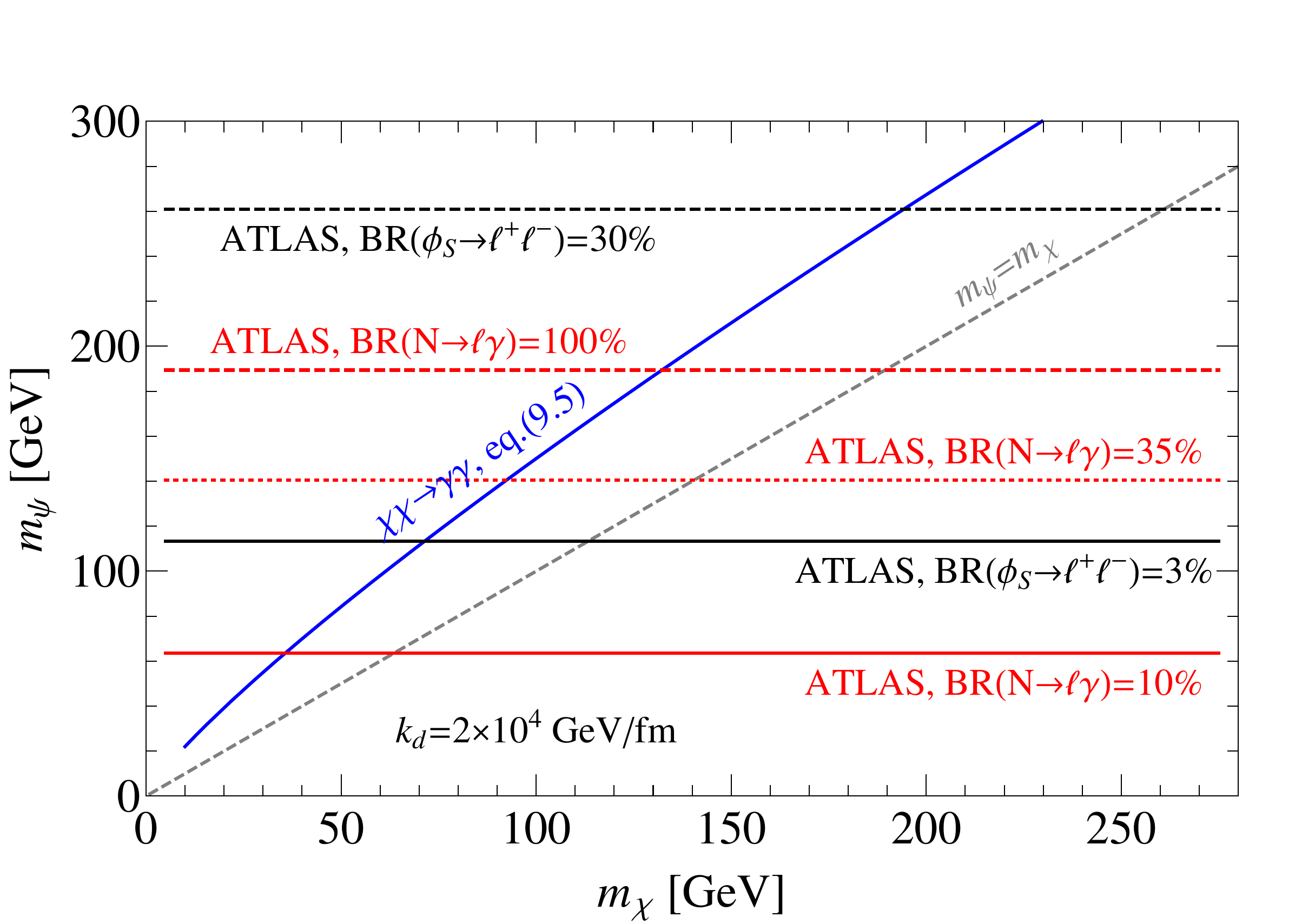}
\caption{
Summary of the LHC constraints on loop model and on MDM with various 
string tensions. 
(a) (upper left): loop model with string tension $k_d=10^4$ GeV/fm; 
(b) (upper right): MDM with string tension $k_d=10^4$ GeV/fm; 
(c) (lower left): MDM with string tension $k_d=5\times 10^3$ GeV/fm; 
(d) (lower right): MDM with string tension $k_d=2 \times 10^4$ GeV/fm. }
\label{fig:summary}
\end{center}
\end{figure}

In case the 130 GeV line is not confirmed in future data, it is
interesting to consider the implications of LHC constraints for
possible future candidates of photophilic dark matter, which could
produce a gamma ray line at some different energy.  To illustrate the
possibilities, we suppose that such a candidate is lurking just beyond
the reach of Fermi's current sensitivity.  The boundary of the 95\%
C.L.\ region found in ref.\ \cite{Fermi-LAT:2013uma} (fig.\ 9, Einasto
profile) is approximately given by
\be
	{\langle\sigma v\rangle_{\rm max}}\cong 0.03\left({m_\chi\over
100{\rm\ GeV}}\right)^{4/3} \langle\sigma v\rangle_0
\ee
(Recall that $\langle\sigma v\rangle_0\equiv 1{\rm\ pb}\cdot c$).
By using this value in eqs.\ (\ref{svl},\,\ref{svm}) rather than the
target value of 0.1 $\langle\sigma v\rangle_0$ that we took in the
case of 130 GeV DM, we obtain constraints on other parameters of 
the models as a function of $m_\chi$, which imply the saturation
of the current Fermi bound on the cross section.  
For the loop model, taking fiducial values $q=2$, $N_c=3$, $\lambda_{\S\chi}=3$
for the parameters other than $m_\S$, 
and recalling that $r\equiv m_\S/m_\chi$,
the constraint reads  
\be
	r^{-4}f(r) = 0.36(m_\chi/100{\rm\ GeV})^{10/3}
\ee 
in the region $m_\chi \lesssim 300$ GeV. The solution of this
transcendental equation is roughly fit by the linear relation
$m_\S = 84.5 + 0.67\, m_\chi$. 
(For higher values of $m_\chi$, the fine-tuned resonance condition
$m_\S\cong m_\chi$ needs to be satisfied to better than 1 part in 100
in order to get such a large cross section, while at $m_\chi\cong
100$ GeV, the tuning is only 50\%.) 
The LHC constraints on the doubly charged 
scalar mass $m_\S$ for the case $k_d=10^4$ GeV/fm are given in the first panel figure of 
fig.\ \ref{fig:summary}. The strongest constraints come from the same-sign dileptons 
in the $e^{\pm} e^{\pm}$ or $\mu^{\pm} \mu^{\pm}$ channel and also  
the opposite-sign dilepton in the s-channel vector boson production. 
From these LHC final states, $m_\S \lesssim (200-270)$ GeV is excluded  
if $m_\S/2 \lesssim m_T \lesssim m_\S$ 
and $\text{BR}(\phi_\S\to\ell^+\ell^-)=30\%$. 
However, these constraints get significantly relaxed if $\eta_{ST}$ decays into $\tau\tau$ dominantly 
and $\phi_\S$ decays into $\eta_S$ or $\eta_T$ dominantly, as discussed before. 
Taking $\text{BR}(\phi_\S\to\ell^+\ell^-)=3\%$, the scalar mass 
$m_\S \lesssim 110$ GeV is excluded, 
which corresponds to dark matter mass around 30 GeV in the loop model.  
The CMS constraints on $\tau$ lepton final states are 
$m_\S \gtrsim (60-80)$ GeV if $m_\S/2 \lesssim m_T \lesssim m_\S$.

Similarly for the magnetic dipole model, in the fiducial case of
maximal mixing ($\cos\theta=1$; see appendix C of \cite{Cline:2012bz}), and taking
$m_\eta\cong\Lambda_d$, we can express $m_\psi$ as a function of 
$m_\chi$ and $\Lambda_d$:  
\be
	m_\psi = m_\chi^{5/6}\,\Lambda_d^{-1/2}
\ee
where all mass scales are in units of 100 GeV. 
The LHC constraints on the fermion mass $m_\psi$ in MDM are shown in 
fig.\ \ref{fig:summary} with three different values of 
the string tension, 
$k_d=(0.5,\, 1,\,2) \times 10^4$ GeV/fm. For the opposite-sign dilepton channel with 
$\text{BR}(\phi_\S\to\ell^+\ell^-)=3\%$, we have $m_\psi \gtrsim (80, 90, 110)$ GeV 
for $k_d=(0.5, 1,2) \times 10^4$ GeV/fm respectively. 
Assuming equal branching ratios into all lepton flavors, 
the excited lepton search respectively constrains 
$m_\psi \gtrsim (170,\, 160,\, 140)$ GeV for the preceding values of 
$k_d$, corresponding  to dark matter masses 
$m_\chi \sim (77,\,88,\,92)$ GeV.

\section{Conclusion}
\label{section:conclusion}

We have presented constraints on two kinds of photophilic dark matter
models that are capable of producing monoenergetic gamma rays in their
annihilations, with energies $\gtrsim 100$ GeV.  Our motivation is
the tentative evidence for 130 GeV gamma rays from the galactic
center, but of course such models could be of interest for future
observations of monoenergetic photons at other energies. Both models require
new charged scalars that are also strongly interacting under an
unbroken SU(N)$_d$ ($d$ for ``dark'') gauge symmetry.  The fact that any such particles
that are pair-produced at LHC must ``hadronize'' to form
SU(N)$_d$-neutral mesonic (or baryonic) states is an essential
feature in determining the constraints on producing such bound
states, either singly or in pairs.  These constraints arise from the
decays of single vector ``mesons'' into lepton pairs, or from decays of
two scalar mesons into two photon pairs. 

A further common feature of the models is that, since any new charged
state must be unstable, there exist couplings that allow them to decay
into right-handed leptons, the simplest possibility amongst standard model 
particles that is allowed
by gauge symmetry.  In one of the models, this results in decays of
charged mesons to like-sign lepton pairs, while in the other the decay
is into a single lepton plus a photon.  Both channels have been
searched for by ATLAS and CMS.

The main conclusion of our study is that features of the models
complementary to the ones most relevant for galactic 130 GeV gamma
rays are constrained by the LHC data.   One of the most important such
parameters is the confinement scale $\Lambda_d$ of the hidden SU(N)$_d$ gauge
group, which sets the scale of the exotic meson masses if $\Lambda_d$ is
greater than the masses of the constituent particles.  In general, we
find that $\Lambda_d$ must not be small compared to the constituent
masses in order to avoid LHC constraints on the meson masses.  For
example, if the vector meson $\phi_\S$ is not far above {310} GeV,
it should be seen in the next run of the LHC.  Since $m_\S = 130-148$
GeV to explain the 130 GeV line in the loop model, $\phi_\S$  should
get a significant fraction of its mass from the dark sector gluons.
However $\Lambda_d$ also cannot be much larger than $m_\S$ in this
model, since the complementary description of the gamma ray line
production in terms of bound state decays into photons, eq.\
(\ref{svs}), shows that some degree of resonant enhancement is required.
In the magnetic dark matter model on the other hand, there is no 
strong prohibition agaisnt taking $\Lambda_d$ large in order to give the
observed 130 GeV line, but an accidental cancellation is needed to 
make $m_\chi=130$ GeV if  $\Lambda_d$ is much greater than $m_\chi$.
We provide a summary of our results (subject to assumptions about
values of other parameters, as discussed in the text) in table
\ref{tab:constraints}.

It is possible that either model will be discovered by exotic
signatures: $\eta_{\S\T} \to \ell^+\ell^+$ (same-sign dileptons) in the loop
model, or $N^-\to \ell^-\gamma$ in the magnetic model.  In the former
case, this depends upon assumptions about the flavor structure of the
dimension-5 operator  $\Lambda_{ij}^{-1} ST^* \bar l_{\R,i}^c
l_{\R,j}$.  The limits or discovery potential are stronger if the
couplings to $e$ or $\mu$ dominate.  Interestingly, the constraint
from $N^-\to \ell^-\gamma$ is less sensitive to hierarchies in the flavor
structure of the relevant coupling $y_i\epsilon_{ab}S^*_a\bar
l_{\R,i}\psi_b$ because the resulting effective interaction is due to
transition magnetic moments that are inversely proportional to the
lepton mass (but proportional to $y_i$).  In the minimally flavor violating case we get a limit of
$m_{N^-}\gtrsim 370$ GeV.

If the 130 GeV line does not persist as data improves, our
results may be of interest in case of future anomalies of this type.
It would be straightforward to generalize the analysis given here for
gamma ray lines at higher energies.  We took a first step in
this direction in section \ref{beyond}.   We believe the models considered
here incorporate several generic features that are useful for
obtaining strong gamma ray lines from dark matter annihilation,
namely, the existence of new charged scalars and a new confining gauge
interaction.

One of the main uncertainties in our treatment was in the prediction
of hadronic matrix elements and bound state masses, involving the
dark SU(N)$_d$ gauge sector.  We used a rather crude model
for predicting these quantities, rescaling with reference to QCD.
If the scenarios presented here become more
motivated by future experimental results, it would be worthwhile to
study these properties within lattice gauge theory to obtain more
accurate predictions.

\begin{table}[t]
\begin{centering}
\vspace{5mm}
\begin{tabular}{|c|c|c|}
\hline
LHC Observable & Constraint & Constraint\\
 & (loop model) & (MD model)\\
\hline
same-sign & BR($\eta_{\S\T}\to ee,\mu\mu)\ll 1$ & $-$\\
 dileptons &	or $m_{\eta_{\S\T}}>200$ GeV   & \\
\hline
vector meson & $m_{\phi_\S} > 310$ GeV &  $m_{\phi_\psi} > 250$ GeV\\
production &  $\Lambda_d >$ few $\times\, m_\S$ &  $\Lambda_d \gtrsim 
300$ GeV\\
\hline
excited lepton searches & $-$ & $m_N > 370$ GeV\\
\hline
diphoton production& $m_{\eta_\S} > 220$ GeV & $m_{\eta_\psi} > 140$ GeV  \\
\hline
4-photon events &  $m_{\eta_\S} > 750$ GeV &  $m_{\eta_\psi} > 600$ GeV  \\
(14 TeV, 100 fb$^{-1}$) & & \\
\hline
monophotons & $-$ &  $-$ \\
\hline
\end{tabular}
\caption{Summary of LHC constraints found for the loop model
and the magnetic dipole model for 130 GeV dark matter.  The
4-photon constraints are projected, based on the ultimate reach of
LHC.}
\label{tab:constraints}
\end{centering}
\end{table}

\bigskip
{\bf Acknowledgments.}  We thank Baris Altunkaynak, David Berge, Ning Chen, 
Guy Moore and Mike Trott for helpful discussions.  JC thanks the
University of Jyv\"askyl\"a Department of Physics for its hospitality
during the completion of this work.  
Our research is supported
by the Natural Sciences and Engineering Research Council (NSERC) of
Canada.

\appendix

\section{Amplitude for photoproduction of $\phi_\S$}
\label{phiSamp}

Section 5.3 of ref.\ \cite{Peskin:1995ev} derives the amplitude for
photoproduction of vector mesons in QCD leading to the result
(\ref{gamma_phi_psi}).  This must be modified for bound states with
orbital angular momentum, for which the wave function vanishes at the
origin.  The general relation for the amplitude of $q\bar q\to
\phi_\S$, from eq.\ (5.44) of \cite{Peskin:1995ev}, is
\be
	{\cal M}(q\bar q\to \phi_\S) = {\sqrt{2 m_{\phi_\S}}\over
	2 E_\S}\int {d^{\,3}k\over (2\pi)^3}\,\tilde\psi^*(\vec k)\,
	{\cal M}(q\bar q\to S S^*)
\label{A1}
\ee
where $\tilde\psi(k)$ is the wave function of $S$ in momentum space.
In the limit where $S$ is nonrelativistic, the matrix element for $q\bar q\to S
S^*$ is given by $(2 e^2 q_q q_\S/ s)\,\bar v_{\bar q}\, 
\vec k\cdot\vec\gamma \,u_q$, so (\ref{A1}) becomes
\be
	{\cal M}(q\bar q\to \phi_\S) = 
	{ e^2 q_q q_\S\sqrt{2 m_{\phi_\S}}\over
	 s\,E_\S}\, \bar v_{\bar q}\, \vec\gamma \,u_q\,
	\cdot \vec\nabla\psi^*(0)
\label{matel}
\ee
Taking the quarks to be relativistic and averaging over their spins
and directions, one finds
\be
	\langle|{\cal M}(q\bar q\to \phi_\S)|^2\rangle = 
	2{e^4 q_q^2 q_\S^2\, m_{\phi_\S}\over 3\, s\, E_\S^2}
	|\vec\nabla\psi(0)|^2
\ee
The same amplitude can be used for $\phi_\S\to e^+ e^-$ by replacing
$q_q \to 1$ and multiplying by $4$ to remove the averaging over
fermion spins.  (There is no need to average over polarizations of
$\phi_\S$ since by holding $\vec\nabla\psi(0)$ fixed we effectively
choose a single polarization.)  Then by computing the decay rate in
the usual way, we arrive at (\ref{gamma_phi_S}).

\section{``Bohr model'' of exotic mesons} 
\label{bohr}
To estimate masses of bound
states in our models, we use a semiclassical quantization approach for
a linear confining potential of the form $V = 2k_d r$, where $k_d$ is
the string tension of the SU(N)$_d$ force, with constituents
whose mass is $m$ and separation is $2r$.  
In the nonrelativistic case, the energy
of the bound state is $p^2/m + V$.  Following Bohr we consider
circular orbits of radius $r$ and demand that the angular momentum
$L = 2pr$ be quantized, $L=n$.
In this way the bound
state energy becomes 
$2m + n^2/(4 r^2 m) + 2k_d r$, which is minimized at
$r_n = (n/2)^{2/3}(m k_d)^{-1/3}$, $p_n = (n k_d m/2)^{1/3}$,
$E_n = 2m + 3(n k_d/2)^{2/3}m^{-1/3}$.
For the relativistic case, the energy becomes $2\sqrt{p^2+m^2} + 2kr$,
again with $2pr=n$, leading to a cubic equation for $p^2$.  If $p\gg
m$ it simplifies to $p=\sqrt{nk/2}$, $E = 2\sqrt{2nk}$.  This model
ignores spin-spin interactions and therefore does not give very accurate
predictions for QCD mesons, but may be more suitable for the
$\eta_\S$ and $\phi_\S$ mesons of the dark matter models, where
$S$ is spin-zero.  In the spirit of the Bohr approach these
are assigned quantum numbers of
$n=1$ and $n=2$ respectively, even though their angular momenta are lower by one
unit.

\end{document}